\documentclass[journal=jctcce,manuscript=article]{achemso}
\usepackage{a4wide}
\usepackage[english]{babel}
\addto\captionsenglish{

}

\usepackage[section]{placeins}   
\usepackage[sc]{mathpazo}        
\usepackage{units}               
\usepackage{xspace}
\usepackage{booktabs}            
\usepackage[table,usenames,dvipsnames]{xcolor}
\usepackage{amsmath}
\usepackage{amssymb}             
\usepackage{multirow}
\usepackage{siunitx}
\usepackage{caption}
\usepackage{circledsteps}        
\usepackage{spreadtab}           
\STsetdecimalsep{.}              
\STautoround{6}                  

\usepackage{todonotes}           

\usepackage{listings}            
\lstdefinelanguage{Dockerfile}
{
  morekeywords={FROM, RUN, CMD, LABEL, MAINTAINER, EXPOSE, ENV, ADD, COPY,
    ENTRYPOINT, VOLUME, USER, WORKDIR, ARG, ONBUILD, STOPSIGNAL, HEALTHCHECK,
    SHELL},
  morecomment=[l]{\#},
  morestring=[b]"
}
\usepackage{tikz}
\usepackage{collcell}
\usepackage{spreadtab}
\usepackage{xstring}

\newcommand*{\MinNumberC}{2.00} 
\newcommand*{\MidNumberC}{5.00} 
\newcommand*{\MaxNumberC}{8.00} 
\newcommand{\ApplyGradientC}[1]{%
    \IfDecimal{#1}{
        \ifdim #1 pt > \MidNumberC pt
            \pgfmathsetmacro{\PercentColor}{max(min(100.0*(#1 - \MidNumberC)/(\MaxNumberC-\MidNumberC),100.0),0.00)} %
            \hspace{-0.33em}\colorbox{green!\PercentColor!yellow}{#1}
        \else
            \pgfmathsetmacro{\PercentColor}{max(min(100.0*(\MidNumberC - #1)/(\MidNumberC-\MinNumberC),100.0),0.00)} %
            \hspace{-0.33em}\colorbox{red!\PercentColor!yellow}{#1}
        \fi
    }{#1}
}
\newcolumntype{C}{>{\collectcell\ApplyGradientC}c<{\endcollectcell}}

\newcommand*{\MinNumberK}{  0.0}
\newcommand*{\MidNumberK}{ 75.0}
\newcommand*{\MaxNumberK}{150.0}
\newcommand{\ApplyGradientK}[1]{%
    \IfDecimal{#1}{
        \ifdim #1 pt > \MidNumberK pt
            \pgfmathsetmacro{\PercentColor}{max(min(100.0*(#1 - \MidNumberK)/(\MaxNumberK-\MidNumberK),100.0),0.00)} %
            \hspace{-0.33em}\colorbox{cyan!\PercentColor!lightgray}{#1}
        \else
            \pgfmathsetmacro{\PercentColor}{max(min(100.0*(\MidNumberK - #1)/(\MidNumberK-\MinNumberK),100.0),0.00)} %
            \hspace{-0.33em}\colorbox{white!\PercentColor!lightgray}{#1}
        \fi
    }{#1}
}
\newcolumntype{K}{>{\collectcell\ApplyGradientK}c<{\endcollectcell}}

\newcommand*{\MinNumber}{0.30} 
\newcommand*{\MidNumber}{2.10} 
\newcommand*{\MaxNumber}{3.90} 
\newcommand{\ApplyGradient}[1]{%
    \IfDecimal{#1}{
        \ifdim #1 pt > \MidNumber pt
            \pgfmathsetmacro{\PercentColor}{max(min(100.0*(#1 - \MidNumber)/(\MaxNumber-\MidNumber),100.0),0.00)} %
            \hspace{-0.33em}\colorbox{green!\PercentColor!yellow}{#1}
        \else
            \pgfmathsetmacro{\PercentColor}{max(min(100.0*(\MidNumber - #1)/(\MidNumber-\MinNumber),100.0),0.00)} %
            \hspace{-0.33em}\colorbox{red!\PercentColor!yellow}{#1}
        \fi
    }{#1}
}
\newcolumntype{R}{>{\collectcell\ApplyGradient}c<{\endcollectcell}}

\newcommand*{\MinNumberX}{  0.0}
\newcommand*{\MidNumberX}{ 57.0}
\newcommand*{\MaxNumberX}{114.0}
\newcommand{\ApplyGradientX}[1]{%
    \IfDecimal{#1}{
        \ifdim #1 pt > \MidNumberX pt
            \pgfmathsetmacro{\PercentColor}{max(min(100.0*(#1 - \MidNumberX)/(\MaxNumberX-\MidNumberX),100.0),0.00)} %
            \hspace{-0.33em}\colorbox{cyan!\PercentColor!lightgray}{#1}
        \else
            \pgfmathsetmacro{\PercentColor}{max(min(100.0*(\MidNumberX - #1)/(\MidNumberX-\MinNumberX),100.0),0.00)} %
            \hspace{-0.33em}\colorbox{white!\PercentColor!lightgray}{#1}
        \fi
    }{#1}
}
\newcolumntype{X}{>{\collectcell\ApplyGradientX}c<{\endcollectcell}}

\newcommand*{\MinNumberQ}{0.160} 
\newcommand*{\MidNumberQ}{1.230} 
\newcommand*{\MaxNumberQ}{2.300} 
\newcommand{\ApplyGradientQ}[1]{%
    \IfDecimal{#1}{
        \ifdim #1 pt > \MidNumberQ pt
            \pgfmathsetmacro{\PercentColor}{max(min(100.0*(#1 - \MidNumberQ)/(\MaxNumberQ-\MidNumberQ),100.0),0.00)} %
            \hspace{-0.33em}\colorbox{green!\PercentColor!yellow}{#1}
        \else
            \pgfmathsetmacro{\PercentColor}{max(min(100.0*(\MidNumberQ - #1)/(\MidNumberQ-\MinNumberQ),100.0),0.00)} %
            \hspace{-0.33em}\colorbox{red!\PercentColor!yellow}{#1}
        \fi
    }{#1}
}
\newcolumntype{Q}{>{\collectcell\ApplyGradientQ}c<{\endcollectcell}}

\newcommand*{\MinNumberY}{ 0.0} 
\newcommand*{\MidNumberY}{35.0} 
\newcommand*{\MaxNumberY}{70.0} 
\newcommand{\ApplyGradientY}[1]{%
    \IfDecimal{#1}{
        \ifdim #1 pt > \MidNumberY pt
            \pgfmathsetmacro{\PercentColor}{max(min(100.0*(#1 - \MidNumberY)/(\MaxNumberY-\MidNumberY),100.0),0.00)} %
            \hspace{-0.33em}\colorbox{cyan!\PercentColor!lightgray}{#1}
        \else
            \pgfmathsetmacro{\PercentColor}{max(min(100.0*(\MidNumberY - #1)/(\MidNumberY-\MinNumberY),100.0),0.00)} %
            \hspace{-0.33em}\colorbox{white!\PercentColor!lightgray}{#1}
        \fi
    }{#1}
}
\newcolumntype{Y}{>{\collectcell\ApplyGradientY}c<{\endcollectcell}}

\newcommand*{\MinNumberP}{0.120} 
\newcommand*{\MidNumberP}{0.880} 
\newcommand*{\MaxNumberP}{1.640} 
\newcommand{\ApplyGradientP}[1]{%
    \IfDecimal{#1}{
        \ifdim #1 pt > \MidNumberP pt
            \pgfmathsetmacro{\PercentColor}{max(min(100.0*(#1 - \MidNumberP)/(\MaxNumberP-\MidNumberP),100.0),0.00)} %
            \hspace{-0.33em}\colorbox{green!\PercentColor!yellow}{#1}
        \else
            \pgfmathsetmacro{\PercentColor}{max(min(100.0*(\MidNumberP - #1)/(\MidNumberP-\MinNumberP),100.0),0.00)} %
            \hspace{-0.33em}\colorbox{red!\PercentColor!yellow}{#1}
        \fi
    }{#1}
}
\newcolumntype{P}{>{\collectcell\ApplyGradientP}c<{\endcollectcell}}

\newcommand*{\MinNumberZ}{ 0.0} 
\newcommand*{\MidNumberZ}{22.0} 
\newcommand*{\MaxNumberZ}{44.0} 
\newcommand{\ApplyGradientZ}[1]{%
    \IfDecimal{#1}{
        \ifdim #1 pt > \MidNumberZ pt
            \pgfmathsetmacro{\PercentColor}{max(min(100.0*(#1 - \MidNumberZ)/(\MaxNumberZ-\MidNumberZ),100.0),0.00)} %
            \hspace{-0.33em}\colorbox{cyan!\PercentColor!lightgray}{#1}
        \else
            \pgfmathsetmacro{\PercentColor}{max(min(100.0*(\MidNumberZ - #1)/(\MidNumberZ-\MinNumberZ),100.0),0.00)} %
            \hspace{-0.33em}\colorbox{white!\PercentColor!lightgray}{#1}
        \fi
    }{#1}
}
\newcolumntype{Z}{>{\collectcell\ApplyGradientZ}c<{\endcollectcell}}

\newcommand*{\MinNumberO}{0.6}%
\newcommand*{\MidNumberO}{1.0} %
\newcommand*{\MaxNumberO}{1.4}%
\newcommand{\ApplyGradientO}[1]{%
    \IfDecimal{#1}{
        \ifdim #1 pt > \MidNumberO pt
            \pgfmathsetmacro{\PercentColor}{max(min(100.0*(#1 - \MidNumberO)/(\MaxNumberO-\MidNumberO),100.0),0.00)} %
            \hspace{-0.33em}\colorbox{green!\PercentColor!yellow}{#1}
        \else
            \pgfmathsetmacro{\PercentColor}{max(min(100.0*(\MidNumberO - #1)/(\MidNumberO-\MinNumberO),100.0),0.00)} %
            \hspace{-0.33em}\colorbox{red!\PercentColor!yellow}{#1}
        \fi
    }{#1}
}
\newcolumntype{O}{>{\collectcell\ApplyGradientO}c<{\endcollectcell}}

\newcommand*{\MinNumberM}{50}%
\newcommand*{\MidNumberM}{84} %
\newcommand*{\MaxNumberM}{118}%
\newcommand{\ApplyGradientM}[1]{%
    \IfDecimal{#1}{
        \ifdim #1 pt > \MidNumberM pt
            \pgfmathsetmacro{\PercentColor}{max(min(100.0*(#1 - \MidNumberM)/(\MaxNumberM-\MidNumberM),100.0),0.00)} %
            \hspace{-0.33em}\colorbox{cyan!\PercentColor!lightgray}{#1}
        \else
            \pgfmathsetmacro{\PercentColor}{max(min(100.0*(\MidNumberM - #1)/(\MidNumberM-\MinNumberM),100.0),0.00)} %
            \hspace{-0.33em}\colorbox{white!\PercentColor!lightgray}{#1}
        \fi
    }{#1}
}
\newcolumntype{M}{>{\collectcell\ApplyGradientM}c<{\endcollectcell}}

\newcommand*{\MinNumberN}{0.43}%
\newcommand*{\MidNumberN}{0.655} %
\newcommand*{\MaxNumberN}{0.88}%
\newcommand{\ApplyGradientN}[1]{%
    \IfDecimal{#1}{
        \ifdim #1 pt > \MidNumberN pt
            \pgfmathsetmacro{\PercentColor}{max(min(100.0*(#1 - \MidNumberN)/(\MaxNumberN-\MidNumberN),100.0),0.00)} %
            \hspace{-0.33em}\colorbox{green!\PercentColor!yellow}{#1}
        \else
            \pgfmathsetmacro{\PercentColor}{max(min(100.0*(\MidNumberN - #1)/(\MidNumberN-\MinNumberN),100.0),0.00)} %
            \hspace{-0.33em}\colorbox{red!\PercentColor!yellow}{#1}
        \fi
    }{#1}
}
\newcolumntype{N}{>{\collectcell\ApplyGradientN}c<{\endcollectcell}}

\newcommand*{\MinNumberL}{3}%
\newcommand*{\MidNumberL}{5.5} %
\newcommand*{\MaxNumberL}{8}%
\newcommand{\ApplyGradientL}[1]{%
    \IfDecimal{#1}{
        \ifdim #1 pt > \MidNumberL pt
            \pgfmathsetmacro{\PercentColor}{max(min(100.0*(#1 - \MidNumberL)/(\MaxNumberL-\MidNumberL),100.0),0.00)} %
            \hspace{-0.33em}\colorbox{cyan!\PercentColor!lightgray}{#1}
        \else
            \pgfmathsetmacro{\PercentColor}{max(min(100.0*(\MidNumberL - #1)/(\MidNumberL-\MinNumberL),100.0),0.00)} %
            \hspace{-0.33em}\colorbox{white!\PercentColor!lightgray}{#1}
        \fi
    }{#1}
}
\newcolumntype{L}{>{\collectcell\ApplyGradientL}c<{\endcollectcell}}

\usepackage[numbers,super,sort&compress]{natbib}
\usepackage{graphicx}            
\usepackage[official]{eurosym}   
\usepackage{url}
\definecolor{LinkCol}{cmyk}{1.00 0.00 0.57 0.30} 

\usepackage[%
colorlinks=true,
linkcolor=LinkCol,
urlcolor=LinkCol,
citecolor=LinkCol]{hyperref}

\newcommand{\gromacs}{GROMACS\xspace}
\newcommand{\nsday}{\nicefrac{ns}{d}\xspace}
\newcommand{\stern}{\mbox{$^\star$}}

\newcommand{\mal}{\mbox{$\times$}\xspace}

\newcommand{\Euro}{\euro\xspace}

\newcommand{\spc}{\hspace{3mm}}

\newcommand{\MPIbpcHG}{Department of Theoretical and Computational Biophysics, 
Max Planck Institute for Multidisciplinary Sciences, 
Am Fassberg 11, 
37077 G\"ottingen, Germany}

\newcommand{\MPIbpcBdG}{Computational Biomolecular Dynamics Group, 
Max Planck Institute for Multidisciplinary Sciences, 
Am Fassberg 11, 
37077 G\"ottingen, Germany}

\newcommand{\AWS}{Amazon Web Services, Amazon Development Center Germany,
Berlin, Germany}

\author{Carsten Kutzner}
\affiliation{\MPIbpcHG}
\email{ckutzne@mpinat.mpg.de}

\author{Christian Kniep}
\affiliation{\AWS}
\email{kniec@amazon.de}

\author{Austin Cherian}
\affiliation{Amazon Web Services Singapore Pte Ltd, 23 Church St, \#10-01, Singapore 049481}

\author{Ludvig Nordstrom}
\affiliation{Amazon Web Services, 60 Holborn Viaduct, London EC1A 2FD, United Kingdom}

\author{Helmut Grubm\"uller}
\affiliation{\MPIbpcHG}

\author{Bert L. de Groot}
\affiliation{\MPIbpcBdG}

\author{Vytautas Gapsys}
\affiliation{\MPIbpcBdG}
\email{vytautas.gapsys@mpinat.mpg.de}

\title{GROMACS in the cloud: A global supercomputer to speed up alchemical drug design}

\begin{document}

\begin{abstract}
We assess costs and efficiency of state-of-the-art high performance cloud computing 
and compare the results to traditional on-premises compute clusters.
Our use case is atomistic simulations carried out with the \gromacs molecular dynamics (MD) toolkit
with a particular focus on the alchemical protein-ligand binding free energy calculations.
Biomolecular simulation is a challenging example of a compute-heavy scientific application
that spans the whole range from high performance computing (HPC) to high throughput computing (HTC),
depending on the questions addressed.
Whereas HPC typically aims at a minimal time-to-solution for a single simulation,
in HTC, the combined output of many independent simulations is maximized.

We set up a compute cluster in the Amazon Web Services (AWS) cloud that incorporates
various different nodes (or instances) with Intel, AMD, and ARM CPUs, some with GPU acceleration.
Using representative biomolecular simulation systems we benchmark how \gromacs performs
on individual instances (for HTC) and across multiple instances (for HPC scenarios).
Thereby we assess which instances deliver the highest performance
and which are the most cost-efficient ones for our use case.

We find that, in terms of total costs including hardware, personnel, room, energy and cooling, 
producing MD trajectories in the cloud can be about as cost-efficient as an on-premises
cluster given that optimal cloud instances are chosen.
Further, we find that high-throughput ligand-screening for protein-ligand binding affinity estimation
can be accelerated dramatically by using global cloud resources.
For a ligand screening study consisting of 19,872 independent simulations
or $\approx 200~\mu{}s$ of combined simulation trajectory,
we used all the hardware that was available in the cloud at the time of the study.
The computations scaled-up to reach peak performances
using more than 4,000 instances, 140,000 cores, and 3,000 GPUs simultaneously around the globe.
Our simulation ensemble finished in about two days in the cloud,
while weeks would be required to complete the task 
on a typical on-premises cluster consisting of several hundred nodes.
We demonstrate that the costs of such and similar studies can be drastically reduced 
with a checkpoint-restart protocol that allows to use cheap Spot pricing and
by using instance types with optimal cost-efficiency.
\end{abstract}

\section{Introduction}
Over the past decades,
molecular dynamics (MD) simulations have become a standard tool to study biomolecules in atomic detail.
In the field of rational drug design, 
MD can greatly reduce costs by transferring parts of the laboratory workflow to the computer.
In the early stage of drug discovery, large libraries of
small molecules with the potential to bind to the target protein (the ``hits'') are identified
and subsequently modified and optimized to ultimately uncover more potent ``lead'' candidates.
\emph{In silico} approaches allow to reduce the number of small molecule compounds from
tens of thousands down to a few hundred entering preclinical studies.

Naturally, it is a combination of all the pharmacokinetic and pharmacodynamic features that defines whether a candidate molecule can be evolved into a useful drug.
Molecular dynamics-based computational drug development concentrates mainly on the particular question of how well a specific ligand binds to a receptor. 
While calculations of absolute protein-ligand binding affinity are feasible, they also present numerous technical challenges.\cite{baumann2021,khalak2021}
Evaluation of the relative binding affinities, however, 
is much more tractable and in the recent years has been well established in the field of computational chemistry.\cite{wang2015,schindler2020,gapsys2020large,kuhn2020}
In the latter approach, MD-based so-called \emph{alchemical} calculations allow obtaining differences in binding free energy between two ligands. 
Such calculations require performing transitions between the two ligands for their protein-bound and for their unbound solvated state.
Carrying out multiple transitions allows to sort the whole collection of ligands by their binding affinity to the target.
Different approaches can be used to carry out the transitions,
but they all involve a $\lambda$ parameter that interpolates between the ligands.
An automated workflow for binding affinity calculations has recently been developed,\cite{gapsys2020large}
based on the open-source software packages pmx\cite{pmx2015} and \gromacs.\cite{pall2020heterogeneous}

Despite continuous advances in hardware and software, 
carrying out MD simulations remains computationally challenging.
A typical MD project can easily occupy a modern compute cluster for days or even months
until a sufficient amount of simulation trajectory is produced.

Where now does a researcher get the required compute time?
Established providers are the compute centers of universities and research institutes, 
national supercomputing centers, and local clusters,
each with particular advantages and disadvantages with respect to 
how easily resources can be accessed, 
how much and how quickly they are available, 
what the costs are, etc.
During the last decade, cloud computing\cite{armbrust2009above,aljamal2019} has developed 
into a new, alternative option to obtain compute time for scientific applications. 

Whereas the first systems of cloud computing reach back into the mid-nineties,\cite{bentley1995}
since about 2007, it is increasingly being used 
for scientific workloads.\cite{garfinkel2007,scientificCloud2008,cloudComp2009,rehr2010scientific, he2010case}
Cloud computing providers like Amazon Web Services (AWS), Microsoft Azure, or Google Cloud Platform
can serve both HPC and HTC demands as they nowadays offer virtually limitless compute power,
plus the possibility to efficiently parallelize individual simulations over multiple instances (compute nodes in the cloud) 
connected by a high-performance network.

One of the main promises of cloud-based computing is its ability to easily scale-up
when the resources for computation are required. 
This way the user has access to an HPC/HTC facility which can flexibly adjust to the particular needs at a given time.
Consequently, the usage of such cloud compute clusters can be fine tuned to optimize costs
or minimize the time-to-solution.

Reports of cloud infrastructure used for MD reach back to 2012. 
Wong et al.\ developed a VMD\cite{humphrey1996vmd} plugin for the NAMD\cite{phillips2005scalable} simulation software 
that simplifies running simulations in the AWS Elastic Compute Cloud (EC2).\cite{wong2012design}
They carried out a simulation of a one million atom large biomolecular system on a total of
64 CPU cores spread over eight EC2 instances.
Van Dijk et al.\ implemented a web portal to execute large-scale parametric MD studies with \gromacs\cite{pall2020heterogeneous}
on European grid resources.\cite{gromacsGrid2012}
In 2014, Kr{\'o}l et al.\ performed an ensemble simulation of 240 replicas of a several hundred atom large evaporating nanodroplet
on 40 EC2 single-core instances.\cite{krol2014}
Kohlhoff et al.\ demonstrated that long simulation timescales for biomolecules can be accessed with the help of cloud computing.
They simulated two milliseconds of dynamics of a major drug target 
on the Google Exacycle platform.\cite{kohlhoff2014cloud}
Singharoy et al.\ described tools to easily perform MD-based flexible fitting of proteins into cryo-EM maps 
in the AWS cloud.\cite{singharoy2016}

The concept of making cloud-based workflows for MD readily available to the scientist is also pursued
by the following projects.
The AceCloud\cite{acecloud2015} on-demand service facilitates running large ensembles of MD simulations in the AWS cloud;
it works with the ACEMD,\cite{acemd2009} \gromacs,
NAMD, and Amber\cite{salomon2013overview} simulation packages.
QwikMD\cite{ribeiro2016qwikmd} is a user-friendly general MD program integrated into VMD and NAMD that
runs on supercomputers or in the AWS cloud.
HTMD\cite{doerr2016htmd} is a python-based extensible toolkit for setting up, running, and analyzing
MD simulations that also comes with an AWS interface.
A purely web-based application that facilitates setting up MD simulations and running them in the cloud 
is described by Nicolas-Barreales et al.\cite{nicolas2021web}
The Copernicus\cite{POUYA2017} scientific computing platform can be used to 
carry out large sets of MD simulations on supercomputers and cloud instances.
In a hands-on fashion, Kohlhoff describes how to perform \gromacs simulations
on Google's cloud platform using Docker containers.\cite{kohlhoff2019google}
Arantes et al.\ propose a Jupyter-notebook based, user-friendly solution to perform
different kinds of MD-related workflows at no cost using the Google Colab services, 
which is especially useful for teaching purposes.\cite{Arantes2021}

Cloud computing has also increasingly being adopted to aid drug discovery.
In their 2013 article,\cite{ebejer2013} Ebejer et al.\ review the use of cloud resources for protein folding and virtual screening
and highlight the potential for future large-scale data-intensive molecular modelling applications.
They point out a virtual screening study of 21 million compounds that has been carried out
in 2012 on a cloud-based cluster with 50,000 cores using Sch{\"o}dinger's docking software Glide.\cite{glide2004}
D'Agostino et al.\ discuss the economic benefits 
of moving \emph{in silico} drug discovery workflows to the cloud.\cite{dAgostino2013cloud}
A recent virtual screening study of 1 billion compounds against proteins involved in \mbox{SARS-CoV-2} infection 
was carried out on Google's cloud services.\cite{gorgulla2021multi}

Cloud computing has been compared to traditional on-premises clusters for exemplary scientific workflows,\cite{montage, costBenefit}
however, we are unaware of a quantitative study so far for the field of MD simulation.
Therefore, here we assess the costs and performance of cloud computing for carrying out biomolecular simulations.
We use \gromacs as the simulation engine and AWS as the provider of cloud infrastructure for this case study,
for the following reasons.
\gromacs is open source and freely available to anyone and it is one of the fastest MD codes available.\cite{gromacs4}
AWS is one of the largest providers of cloud infrastructure, on par with Microsoft and Google.\cite{aljamal2019}

First, we measured the \gromacs performance using established benchmarks\cite{kutznerMoreBang2018}
on a broad range of available instance types (with and without GPUs), and also across multiple instances.
The simulation performance to instance price ratio allows to 
optimize for a minimal time-to-solution or minimal project costs.
The benchmark results and the instance costs allowed us to compare
the costs of carrying out simulations in the cloud to those for operating an in-house cluster.
Second, we ask how much high throughput ligand screening can be accelerated in the cloud.
To address this question, 
we used globally available compute capacity to carry out a large protein-ligand binding affinity study
at highest possible throughput.

\section{General background}
\subsection{Cloud computing}
The large cloud providers offer a wide range of instance types, with and without GPUs,
optionally with extra memory or HPC network, targeted towards different application areas.
The compute unit that is rented out to customers is called \emph{instance}.
It may be a whole node with multiple cores and GPU(s),
or just a part of a node, even just a single core.

Large nodes that are rented out as several smaller instances 
are shared between different customers.
However, each customer is restricted to her instance (her part of the node) exclusively,
and her processes cannot spill over into the compute cores, memory, or network bandwidth allocated to other instances on the node.
AWS instances come with a certain number of virtual CPUs (vCPUs)
which translate to hardware threads.
Renting two vCPUs on a modern AMD or Intel-based instance
is equivalent to getting one physical core exclusively on that machine.

Although the actual exact location of allocated compute instances remains opaque to the user,
the \emph{region} she chooses encompasses a group of geographically close data centers.
Costs usually vary by region, depending on supply and demand, as well as energy costs, and
specific services or cutting edge processor features may only be available in some of the regions.
For the case of AWS, each region consists of multiple, isolated, and physically separate \emph{availability zones} (AZs) within a geographic area.
An AZ is a group of one or more datacenters 
with independent redundant power supply and network connectivity. 
In 2021, AWS had 85 AZs in 26 regions.

There are different payment models that can be chosen from.
\emph{On-demand} payment is most flexible,
as one can rent an instance at any time and give it back when it is not needed any more. 
One only pays for the time that the instance is needed.
One can also get \emph{reserved instances} at a 50--70\% discount if one books these instances for one to three years,
but then one has to pay regardless if one can make use of them.
\emph{Preemptible} or \emph{Spot} instances tap into the pool of currently unused compute capacity
and are available at discount rates of up to 90\% compared to on-demand,
though pricing varies across AZs and over time.
However, a Spot instance can be claimed back at any time by Amazon EC2 with a two-minute warning.

\subsection{Using hardware efficiently with \gromacs}
\label{sec:gromacs}
Key to optimal simulation performance is understanding how \gromacs makes use of the available hardware.
\gromacs combines several parallelization techniques,
among them MPI and OpenMP parallelism, GPU offloading,
and separable ranks to evaluate long-range electrostatics.
With domain decomposition (DD),
the simulation system is divided into $n_x \times n_y \times n_z$ domains,
each of which is operated on by one MPI rank.\cite{gromacs4}
During the simulation, dynamic load balancing (DLB) adjusts the size of the domains 
such that any uneven computational load between the MPI ranks is minimized.

Each MPI rank can further have multiple OpenMP threads.
Best performance is usually achieved when the product of MPI ranks and OpenMP threads
equals the number of cores (or hardware threads) on a node or instance,
and when all threads are properly pinned to cores.
Though leaving some cores idle may in rare cases make sense,
oversubscription will lead to significant performance degradation.

When distributing a simulation system over an increasing number of MPI ranks in a strong scaling scenario, at some point the time spent for communication between the ranks limits further speedup.
Usually the bottleneck is in the long-range contribution to the electrostatic forces
which are calculated with the Particle Mesh Ewald (PME) method.\cite{Essmann1995}
Parallel PME requires all-to-all communication between the participating ranks,
leading to $r^2$ MPI messages being sent on $r$ MPI ranks.\cite{gromacs4}
This communication bottleneck can be alleviated by assigning a subset of MPI ranks
to exclusively evaluate the long range PME part.
As typically only a quarter up to a third of all ranks need to be allocated for long range electrostatics, 
the communication bottleneck is greatly reduced, yielding better performance and scalability.

\gromacs can offload various types of computationally demanding interactions onto the GPU.\cite{pall2014,abraham2015,kutznerMoreBang2018}
One of the largest performance benefits stems from offloading the short-range part of the nonbonded interactions
(Coulomb and van der Waals).
In parallel, each MPI rank can offload its local domain's interactions to a GPU.
The PME long range part can be offloaded as well, 
however, this computation still cannot be distributed onto multiple GPUs. 
Additionally, bonded interactions and for suitable parameter settings the integration and constraint calculations can be offloaded.

The relative GPU to CPU compute power on a node determines how many interaction types
can be offloaded for optimal performance.
Ideally, CPU and GPU finish their force calculation at about the same time in the MD time step
so that no time is lost waiting.

Earlier studies showed that both the \gromacs performance as well as the performance to price (P/P) ratio, 
i.e.\ how much MD trajectory is produced per invested \Euro,
can vastly differ for different hardware.\cite{kutznerBestBang2015,kutznerMoreBang2018}
Nodes with GPUs provide the highest single-node \gromacs performance.
At the same time, P/P skyrockets when consumer GPUs are used instead of
professional GPUs (e.g. NVIDIA GeForce RTX instead of Tesla GPUs). 
The P/P ratio of consumer GPU nodes is typically at least a factor of three higher
than that of CPU nodes or nodes with professional GPUs.

Pronounced variations in \gromacs performance and cost-efficiency are therefore expected
between the different instance types on AWS.
Benchmarks allow to pick instance types optimal for MD simulation.

\subsection{Obtaining relative binding free energies from MD simulations}
To evaluate relative binding affinities in a chemical library of interest,
ligands are connected into a network (graph) and a number of pair-wise calculations is performed
eventually allowing to sort the molecules according to their binding free energy.
It is a usual practice to repeat calculations several times for each ligand pair to obtain reliable uncertainty estimates.\cite{knapp2018,gapsys2020elife,bhati2018uncertainty}

Various methods for the alchemical calculations have been developed.
For example, the commercially available Schr{\"o}dinger software uses a free energy perturbation based approach,\cite{Wang2019}
whereas the open source workflow used here\cite{pmx2015, gapsys2020large} is based on thermodynamic integration (TI)\cite{vanGunsteren1993TI}
using a non-equilibrium transition protocol.\cite{gapsys2015calculation}
Both approaches yield similarly accurate relative binding free energies at similar computational effort.\cite{gapsys2020large}

The non-equilibrium TI approach requires equilibrated ensembles of the physical end states for the solvated protein with ligand,
one for ligand A and one for ligand B, as well as two equilibrated ensembles of ligand A and ligand B in solution.
From the equilibrated ensembles, many short "fast growth" TI simulations are spawned
during which ligand A is transformed into ligand B and vice versa using a $\lambda$-dependent Hamiltonian.
\label{sec:introTI}
The free energy difference is then derived from the overlap of the forward (A $\rightarrow$ B)
and reverse (B $\rightarrow$ A) work distributions using estimators based on the Crooks Fluctuation theorem.\cite{Crooks1999}


\section{Methods}
We will first describe the setup of the cloud-based HPC clusters that
we used to derive the \gromacs performance on a range of available instance types,
and provide some details about the benchmark input systems
and on how the benchmarks were carried out.
Then we will outline our setup to distribute a large ensemble of free energy calculations
on globally available compute resources.

\subsection{Cloud-based HPC cluster and software setup}
The benchmark simulations were carried out on
AWS compute clusters in the North Virginia region
set up with the ParallelCluster\cite{parallelCluster} open source cluster management tool.
Each cluster consists of a master instance of the same architecture as the nodes (x86 or ARM).
The master fires up and closes down the node instances as needed 
and operates the queueing system (SLURM).\cite{slurm2003}
For the x86 cluster, we used ParallelCluster v.\ 2.10.0 on a \texttt{c5.2xlarge} master, 
for the ARM cluster, we used v.\ 2.9.1 on a \texttt{m6g.medium} master instance.
For brevity, we will from now on refer to \texttt{c5.2xlarge} instances as \texttt{c5.2xl}
and also abbreviate all other \texttt{*xlarge} instances accordingly.
All instances use Amazon Linux 2 as operating system,
for technical specifications of the instances see Tab.~\ref{tab:instances}.
Whereas all instances can communicate via TCP (see last columns in Tab.~\ref{tab:instances} for the network bandwidth),
some of them have an Elastic Fabric Adapter (EFA).
EFA enables HPC scalability across instances by a higher throughput compared to TCP
and a lower and more consistent latency.

\begin{table}
\caption{Technical specifications of AWS instances used in this study, 
and \gromacs compilation options.
i = using Intel MPI 2019, t = using \gromacs' built-in thread-MPI library.
EFA (Elastic Fabric Adapter) signals whether an HPC network is available.}
\label{tab:instances}
\footnotesize
\begin{tabular}{llcclccrc}
\toprule
instance           & CPU                   & HT or & clock & used SIMD                & NVIDIA      & MPI & \multicolumn{2}{c}{network}       \\
type               & model                 & vCPUs & (GHz) & instructions             & GPUs        & lib & (Gbps)               & EFA        \\ \midrule
\texttt{c5.24xl}   & Intel 8275CL          &   96  &  3.0  & AVX\_512                 &   --        &  i  &     25               &            \\
\texttt{c5.18xl}   & Intel 8124M           &   72  &  3.0  & AVX\_512                 &   --        &  i  &     25               &            \\
\texttt{c5n.18xl}  & Intel 8124M           &   72  &  3.0  & AVX\_512                 &   --        &  i  &    100               & \checkmark \\
\texttt{c5.12xl}   & Intel 8275CL          &   48  &  3.0  & AVX\_512                 &   --        &  i  &     12               &            \\
\texttt{c5.9xl}    & Intel 8124M           &   36  &  3.0  & AVX\_512                 &   --        &  i  &     10               &            \\
\texttt{c5.4xl}    & Intel 8275CL          &   16  &  3.0  & AVX\_512                 &   --        &  i  & $\le$ 10             &            \\
\texttt{c5.2xl}    & Intel 8275CL          &    8  &  3.0  & AVX\_512                 &   --        &  i  & $\le$ 10             &            \\
\texttt{c5.xl}     & Intel 8275CL          &    4  &  3.0  & AVX\_512                 &   --        &  i  & $\le$ 10             &            \\
\texttt{c5.large}  & Intel 8124M           &    2  &  3.0  & AVX\_512                 &   --        &  i  & $\le$ 10             &            \\\midrule
\texttt{c5a.24xl}  & AMD EPYC 7R32         &   96  &  3.3  & AVX2\_128                &   --        &  i  &     20               &            \\
\texttt{c5a.16xl}  & AMD EPYC 7R32         &   64  &  3.3  & AVX2\_128                &   --        &  i  &     20               &            \\
\texttt{c5a.12xl}  & AMD EPYC 7R32         &   48  &  3.3  & AVX2\_128                &   --        &  i  &     12               &            \\
\texttt{c5a.8xl}   & AMD EPYC 7R32         &   32  &  3.3  & AVX2\_128                &   --        &  i  &     10               &            \\
\texttt{c5a.4xl}   & AMD EPYC 7R32         &   16  &  3.3  & AVX2\_128                &   --        &  i  & $\le$ 10             &            \\
\texttt{c5a.2xl}   & AMD EPYC 7R32         &    8  &  3.3  & AVX2\_128                &   --        &  i  & $\le$ 10             &            \\
\texttt{c5a.xl}    & AMD EPYC 7R32         &    4  &  3.3  & AVX2\_128                &   --        &  i  & $\le$ 10             &            \\
\texttt{c5a.large} & AMD EPYC 7R32         &    2  &  3.3  & AVX2\_128                &   --        &  i  & $\le$ 10             &            \\\midrule
\texttt{c6g.16xl}  & ARM Graviton2         &   64  &  2.3  & NEON\_ASIMD\hspace{-6mm} &   --        &  t  &     25               &            \\
\texttt{c6g.12xl}  & ARM Graviton2         &   48  &  2.3  & NEON\_ASIMD\hspace{-6mm} &   --        &  t  &     20               &            \\
\texttt{c6g.8xl}   & ARM Graviton2         &   32  &  2.3  & NEON\_ASIMD\hspace{-6mm} &   --        &  t  & $\le$ 10             &            \\
\texttt{c6g.4xl}   & ARM Graviton2         &   16  &  2.3  & NEON\_ASIMD\hspace{-6mm} &   --        &  t  & $\le$ 10             &            \\
\texttt{c6g.2xl}   & ARM Graviton2         &    8  &  2.3  & NEON\_ASIMD\hspace{-6mm} &   --        &  t  & $\le$ 10             &            \\
\texttt{c6g.xl}    & ARM Graviton2         &    4  &  2.3  & NEON\_ASIMD\hspace{-6mm} &   --        &  t  & $\le$ 10             &            \\\midrule
\texttt{c6i.32xl}  & Intel 8375C           &  128  &  2.9  & AVX\_512                 &   --        &  i  &     50               & \checkmark \\
\texttt{m6i.32xl}  & Intel 8375C           &  128  &  2.9  & AVX\_512                 &   --        &  t  &     50               & \checkmark \\\midrule
\texttt{m5n.24xl}  & Intel 8259CL          &   96  &  2.5  & AVX\_512                 &   --        &  i  &    100               & \checkmark \\                     
\texttt{m5zn.12xl} & Intel 8252C           &   48  &  3.8  & AVX\_512                 &   --        &  t  &    100               & \checkmark \\
\texttt{m5zn.2xl}  & Intel 8252C           &    8  &  3.8  & AVX\_512                 &   --        &  t  & $\le$ 25             &            \\\midrule
\texttt{p3.2xl}    & Intel E5-2686v4       &    8  &  2.3  & AVX2\_256                & V100        &  t  & $\le$ 10             &            \\
\texttt{p3.8xl}    & Intel E5-2686v4       &   32  &  2.3  & AVX2\_256                & V100 \mal 4 &  t  &     10               &            \\
\texttt{p3.16xl}   & Intel E5-2686v4       &   64  &  2.3  & AVX2\_256                & V100 \mal 8 &  t  &     25               &            \\
\texttt{p3dn.24xl} & Intel 8175M           &   96  &  2.5  & AVX2\_256                & V100 \mal 8 &  t  &    100               & \checkmark \\
\texttt{p4d.24xl}  & Intel 8275CL          &   96  &  3.0  & AVX2\_256                & A100 \mal 8 &  i  &    400               & \checkmark \\
\texttt{g3s.xl}    & Intel E5-2686v4       &    4  &  2.3  & AVX2\_256                & M60         &  i  &     10               &            \\
\texttt{g3.4xl}    & Intel E5-2686v4       &   16  &  2.3  & AVX2\_256                & M60         &  i  & $\le$ 10             &            \\
\texttt{g4dn.xl}   & Intel 8259CL          &    4  &  2.5  & AVX\_512                 & T4          &  i  & $\le$ 10             &            \\
\texttt{g4dn.2xl}  & Intel 8259CL          &    8  &  2.5  & AVX\_512                 & T4          &  i  & $\le$ 25             &            \\
\texttt{g4dn.4xl}  & Intel 8259CL          &   16  &  2.5  & AVX\_512                 & T4          &  i  & $\le$ 10             &            \\
\texttt{g4dn.8xl}  & Intel 8259CL          &   32  &  2.5  & AVX\_512                 & T4          &  i  &     50               &            \\
\texttt{g4dn.12xl} & Intel 8259CL          &   48  &  2.5  & AVX\_512                 & T4          &  i  &     50               &            \\
\texttt{g4dn.16xl} & Intel 8259CL          &   64  &  2.5  & AVX\_512                 & T4          &  i  &     50               &            \\
\texttt{g4dn.12xl} & Intel 8259CL          &   48  &  2.5  & AVX\_512                 & T4 \mal 4   &  i  &     50               &            \\
\bottomrule
\end{tabular}
\end{table}

Different versions of \gromacs (2020.2 and 2021.1, with and without MPI) 
were installed with the Spack\cite{spack} 0.15.4 package manager.
\gromacs was built in mixed precision with GCC 7.3.1, FFTW 3.3.8, hwloc 1.11, and either
Intel MPI 2019 or it's built-in thread-MPI library (as listed in Tab.~\ref{tab:instances}).
GPU versions used CUDA 10.2 on \texttt{g} instances and CUDA 11.1 on \texttt{p} instances.
Benchmarks on \texttt{m6i.32xl} instances were done using ICC 2021.2 and Intel MKL.
The multi-instance scaling benchmarks on \texttt{m5n.24xl} and \texttt{m5zn.12xl} instances 
used a \gromacs executable built with Intel MPI + ICC 2021.2 and Intel MKL.

A workshop to reproduce a (slightly updated) setup is available on the web,\cite{gromacsPcluster2021}
whereas general advice on how to use AWS services can be found in this book.\cite{wittig2018amazon} 

\subsection{Description of the MD benchmark systems}
To determine the \gromacs performance on various instance types
we used seven simulation systems (Tab.~\ref{tab:systems}).
MEM, RIB, and PEP are typical MD systems differing in size and composition, 
where no special functionality like external forces or free energy (FE) is required.
MEM is an aquaporin tetramer embedded in a lipid membrane surrounded by water and ions
in a simulation box of $10.8 \times 10.2 \times 9.6~\mbox{nm}^3$ size.\cite{escidoc:599912}
RIB contains an \emph{E. coli} ribosome in a box of size (31.2 nm)$^3$ with water and ions.\cite{escidoc:1854600}
The (50 nm)$^3$ large PEP system was used to study peptide aggregation;\cite{Matthes2012}
it contains 250 steric zipper peptides in solution.
MEM, RIB, and PEP were used in previous performance studies,\cite{kutznerBestBang2015, kutznerMoreBang2018, Kutzner2014}
allowing to compare cloud instances to a variety of other already benchmarked hardware. 

\begin{table}
\begin{center}
\caption{{\bf Benchmark systems.}
Specifications of the MD input systems that are used for benchmarks in this study.
FE column lists the number of perturbed atoms for this benchmark
(note that this number will vary for different ligands considered in the physical end states),
$\Delta t$ is integration time step, $r_c$ cutoff radius, grid sp. the spacing of the PME grid.
Benchmark input \texttt{.tpr} files can be downloaded from \url{https://www.mpinat.mpg.de/grubmueller/bench}}
\label{tab:systems}
\small
\begin{tabular}{lrcccr} \toprule
benchmark                      &     \# of & $\Delta t$ & $r_c$ & grid sp.& \# of FE\\
acronym                        &   atoms   & (fs)       & (nm)  & (nm)    & atoms   \\ \midrule
PEP\cite{Kutzner2014}          &12,495,503 &     2      &  1.2  &   0.160 &   0     \\
RIB\cite{escidoc:1854600}      & 2,136,412 &     4      &  1.0  &   0.135 &   0     \\
MEM\cite{escidoc:599912}       &    81,743 &     2      &  1.0  &   0.12  &   0     \\ \midrule
SHP-2 protein  + ligand        &   107,330 &     2      &  1.1  &   0.12  &  53     \\
c-Met protein  + ligand        &    67,291 &     2      &  1.1  &   0.12  &  61     \\
HIF-2$\alpha$ protein + ligand &    35,546 &     2      &  1.1  &   0.12  &  35     \\
c-Met ligand in water          &     6,443 &     2      &  1.1  &   0.12  &  61     \\
\bottomrule
\end{tabular}
\end{center}
\end{table}

c-Met, HIF-2$\alpha$, and SHP-2 are representative systems from the large binding affinity ensemble assembled by Schindler et al.~\cite{schindler2020}
These systems run special FE kernels 
for all $\lambda$-dependent interactions, i.e.\ those involving a transition between atomic properties.
As the FE kernels are slower than the normal kernels,
and due to a larger cutoff, finer PME grid and the need to calculate two PME grids (one for each of the physical states),
even at equal atom count a FE simulation will be slower than a plain MD system.
We therefore carried out separate benchmarks for the FE systems,
chosen such that predicting the performance of all ensemble members listed in Tabs.~\ref{tab:ba1}--\ref{tab:ba2}
is easily possible: A small, medium and large protein plus ligand system to cover the whole
range of sizes for the protein systems (35~k -- 110~k atoms) and one ligand in water system
representative for all 9,936 ligand in water systems. 

\begin{table}[tb]
\caption{{\bf Systems considered for the first binding affinity study.} 
For each of eight considered protein-ligand complexes (from the study~\cite{schindler2020}) two sets of simulations are performed: 
\emph{protein+ligand} for the solvated protein-ligand complex and \emph{ligand} for the solvated ligand alone.
An \emph{edge} is referred to as the transition of one ligand A to another ligand B.
As we probe three independent replicas for each system in forward and backward simulation direction,
and three small molecule force fields (GAFF~\cite{wang2004} v2.11, CGenFF~\cite{vanommeslaeghe2010cgenff,yesselman2012} v3.0.1 and OpenFF~\cite{qiu2020} v2.0.0),
the total number of jobs is $3 \times 2 \times 3 = 18 \times$ the number of edges for the \emph{protein+ligand} 
plus an equal number for the \emph{ligand} systems.
}
\label{tab:ba1}
\begin{center}
\begin{tabular}{lrrrr}
\toprule
                & \multicolumn{2}{c}{size (atoms)} & \# of    & \# of              \\
system          & protein+ligand &      ligand     & edges    & jobs               \\ \midrule
CDK8            &       109,807  &       5,789     &    54    &             972    \\
SHP-2           &       107,330  &       6,231     &    56    &           1,008    \\
PFKFB3          &        96,049  &       6,570     &    67    &           1,206    \\
Eg5             &        79,653  &       6,116     &    65    &           1,170    \\
c-Met           &        67,291  &       6,443     &    57    &           1,026    \\
SYK             &        66,184  &       5,963     &   101    &           1,818    \\
TNKS2           &        52,251  &       6,012     &    60    &           1,080    \\
HIF-2$\alpha$   &        35,546  &       4,959     &    92    &           1,656    \\ \midrule
Total           &                &                 &          & 2$\times$ 9,936    \\
\bottomrule
\end{tabular}
\end{center}
\end{table}

In total, $2 \times 9,936 = 19,872$ independent jobs were run for the binding affinity study (Tab.~\ref{tab:ba1})
by which 1,656 free energy differences ($\Delta\Delta G$ values) were determined.
Each job first simulated six nanoseconds at equilibrium (for the starting state, i.e.\ A or B),
followed by 80 non-equilibrium transitions from the start to the end state (A $\rightarrow$ B, or B $\rightarrow$ A),
as mentioned in section \ref{sec:introTI}.
The 80 individual transitions were started from different, equidistant, positions of the equilibrium trajectory
and were each 50 picoseconds long. 
In total 10 nanoseconds of trajectory were generated per job.

\begin{table}[tb]
\caption{{\bf Systems considered for the second binding affinity study.}
Same as in Tab.~\ref{tab:ba1}, but considering 14 protein-ligand complexes in one MD force field (OpenFF v2.0.0).
The systems were collected from public sources for the previous free energy calculation studies~\cite{perez2019predicting,gapsys2020large}.
The total number of jobs is $3 \times 2 = 6 \times$ the number of edges for the \emph{protein+ligand} 
plus an equal number for the \emph{ligand} systems.
}
\label{tab:ba2}
\begin{center}
\begin{tabular}{lrrrr}
\toprule
                & \multicolumn{2}{c}{size (atoms)} & \# of    & \# of              \\
system          & protein+ligand &      ligand     & edges    & jobs               \\ \midrule
CDK2            &       106,910  &       4,993     &    25    &             150    \\
P38             &        80,777  &       6,750     &    56    &             336    \\
ROS1            &        73,957  &       8,434     &    63    &             378    \\
Bace            &        73,330  &       5,914     &    58    &             348    \\
JNK1            &        72,959  &       5,956     &    31    &             186    \\
Bace (Hunt)     &        72,036  &       5,773     &    60    &             360    \\
Bace (p2)       &        71,671  &       6,687     &    26    &             156    \\
PTP1B           &        70,020  &       8,753     &    49    &             294    \\
PDE2            &        63,943  &       5,504     &    34    &             204    \\
TYK2            &        62,292  &       5,956     &    24    &             144    \\
PDE10           &        56,616  &       7,655     &    62    &             372    \\
Thrombin        &        49,312  &       6,025     &    16    &              96    \\
Galectin        &        35,635  &       9,576     &     7    &              42    \\
MCL1            &        32,745  &       5,435     &    71    &             426    \\ \midrule
Total           &                &                 &          & 2$\times$ 3,492    \\
\bottomrule
\end{tabular}
\end{center}
\end{table}

\subsection{Benchmarking procedure}
\subsubsection{MEM and RIB plain MD systems}
MEM and RIB benchmarks were run for 20~k steps on single instances and for 40~k -- 50~k steps
when scaling across multiple instances or multiple GPUs using \gromacs 2020.
Due to effects of load balancing, PME grid versus cutoff scaling and memory allocations (compare section~\ref{sec:gromacs})
the first few thousand steps in a \gromacs simulation are typically slower than average
and were therefore excluded from the benchmarks,
which are intended to be proxies for the long-term performance.

To make use of all CPU cores of an instance, 
the product of ranks $\times$ threads was set to the number of physical cores
or to the number of available hardware threads.
We benchmarked various combinations of ranks $\times$ threads and 
additionally checked whether separate PME ranks improve performance.
Pinning of threads to cores was enabled, and no checkpoint files were written during the benchmark runs.

On GPU instances we used one rank per GPU and offloaded all short range nonbonbed interactions to the GPU(s).
For improved performance, also the long range PME contribution was offloaded to a GPU,
except for some GPU instances with many cores,
where it turned out to be faster to evaluate the long range PME contribution on the CPU.
For scaling benchmarks across two or more GPU instances,
the long range PME contribution was run on the CPU part,
as only there it can be parallelized.

The timings (in simulated nanoseconds per day)
reported for MEM and RIB (Tabs.~\ref{tab:numbers2020}--\ref{tab:g4dn.16xl_scaling2020})
are averages over two runs.
The parallel efficiency on $n$ instances $E_n$ reported in Tabs.~\ref{tab:c5n_scaling2020}, \ref{tab:g4dn.8xl_scaling2020}, and \ref{tab:g4dn.16xl_scaling2020}
is computed as the performance $P_n$ on $n$ instances divided by $n$ times the performance on a single instance:
\begin{equation}
E_n = \frac{P_n}{n \cdot P_1}
\end{equation}

The performance to price ratios (ns/\$) in the MEM and RIB tables are calculated from
Amazon EC2 on-demand prices for Linux instance in the US East (N. Virginia) region
(\url{https://aws.amazon.com/ec2/pricing/on-demand/}).

\subsubsection{Free energy systems used for the binding affinity study}
\label{sec:methodsFEbench}
Each job of the binding affinity ensemble run (Tab.~\ref{tab:systems}) consists of two parts,
first, a 6~ns equilibrium simulation, 
second, 80 non-equilibrium transitions of 50~ps length each.

The first (equilibration) part was benchmarked as described above for MEM and RIB,
using 10~k total steps with timings from the first half discarded.
In cases, where PME grid vs.\ cutoff tuning took more than 5~k time steps, 
20~k time steps were used in total.
For the binding affinity runs we did not check whether separate PME ranks improve the performance.
The timings reported in tables \ref{tab:equil_CPU}--\ref{tab:equil_GPU}
resulted from individual runs of the equilibration part.
Here, we ran on individual instances only,
no scaling across multiple instances was attempted.
Though in most cases we tested various combinations 
of splitting a given number of total cores $N_\text{c}$ into ranks and threads 
$N_\text{c} = N_\text{ranks} \times N_\text{threads}$, 
we do not report all results 
in tables \ref{tab:equil_CPU}--\ref{tab:equil_GPU} 
to keep them reasonably concise.
Instead, we report a consensus for the combination $N_\text{ranks} \times N_\text{threads}$ that
yielded best results across the free energy benchmark systems.

The second (transition) part was benchmarked by timing
one of the 50~ps (25~k steps) long transition runs.
No time steps were discarded from the measurements,
as an initial below average performance will occur in each of the 80 short transition runs
and thus should be included when predicting the performance of the whole transition part.

The total costs per free energy difference (Fig.~\ref{fig:optimalConfig})
have been derived by combining the equilibration and transition phase timings 
of the protein-ligand complex and the ligand alone in water.
Six runs were performed per free energy difference for the protein-ligand complex
(3 replicas $\times$ 2 directions) plus additional six for the solvated ligand.
All runs for the solvated ligand were performed on \texttt{c5.2xl} instances. 
On-demand (or Spot) prices for AWS instances in the US East (N. Virginia) region as of May 2021 were used.

\subsection{Setup of globally distributed compute resources}
\label{sec:hyperbatch}

\begin{figure}[tb]
\begin{center}
\includegraphics[width=\textwidth]{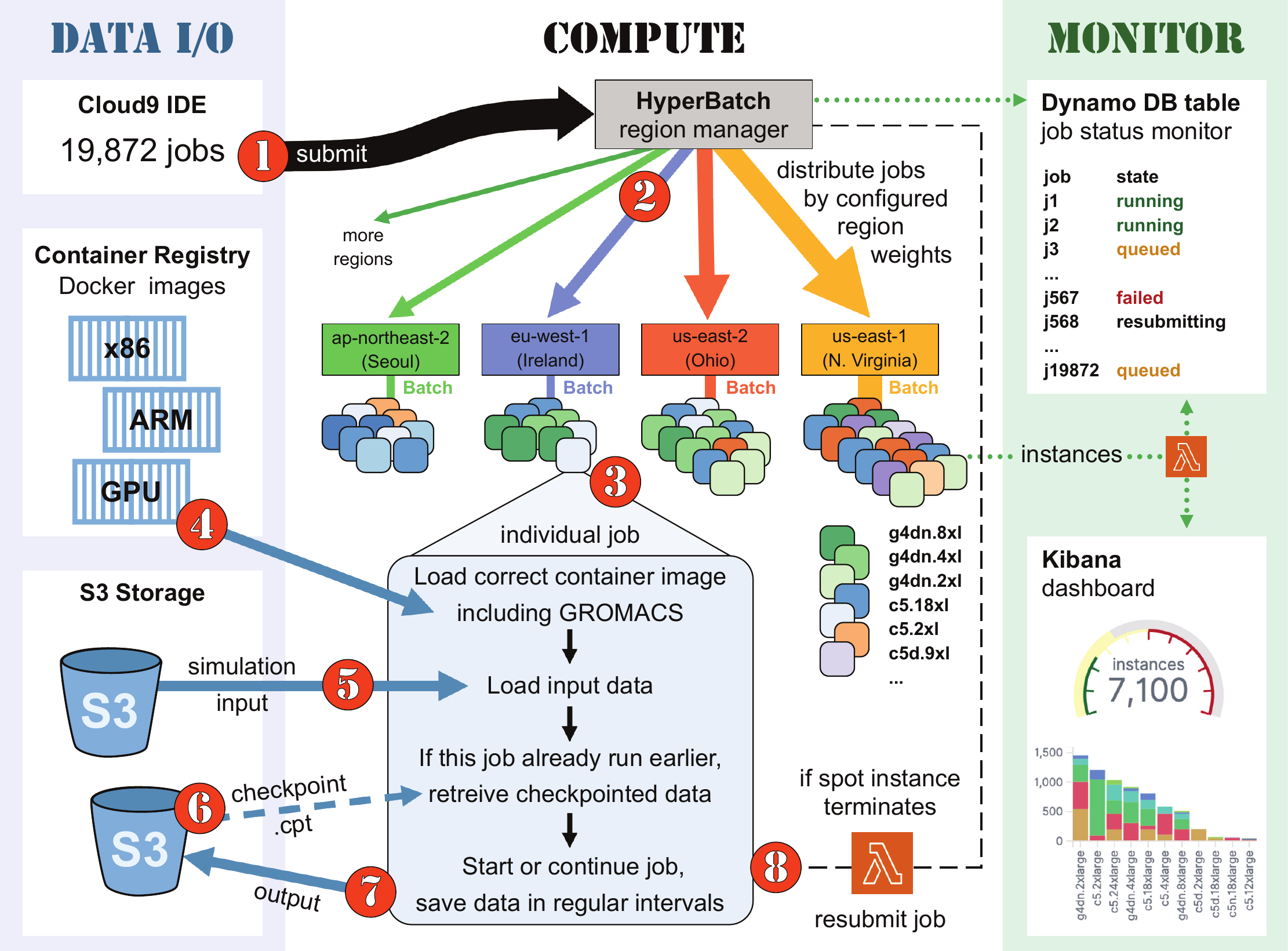}
\caption{{\bf The HyperBatch-based setup distributes all 19,872 \gromacs jobs globally.}
An illustrative lifetime of a job follows the steps \Circled{1},\ldots \Circled{8} 
and is described in Section~\ref{sec:hyperbatch} of the text.}
\label{fig:batchworkflow}
\end{center}
\end{figure}

The allocation of cloud-based compute resources (e.g.\ via ParallelCluster or AWS Batch\cite{AwsBatch}) 
is normally confined to a specific geographic region (there are currently 26 in AWS).
Whereas stacks of small to medium jobs can be conveniently executed using just a single region, 
a global setup is better suited when a substantial amount of core hours is needed:
The pool of available instances is much larger for all regions combined
compared to just a single region. This allows, e.g., to start more instances at the same time,
or to pick only the subset of instances with the highest performance to price ratio.
To benefit from global compute resources, we used AWS HyperBatch
as a means to provide a single entry point for jobs scheduled to AWS Batch
queues across regions.
 
The technical setup used for the binding affinity study
is sketched in Fig.~\ref{fig:batchworkflow}.
For easier overview, the main compute setup is shown in the middle, 
whereas input and output of data is gathered in the left, blue column, and
monitoring functionality about the status of jobs and instances in the right, green column.
In a nutshell, AWS HyperBatch provides cross-regional serverless job scheduling and
resource orchestration using DynamoDB, Lambda functions, Step Functions,
AWS Kinesis Data Streams, the Simple Queue Service (SQS), and the Amazon API Gateway.\cite{wittig2018amazon}

For the binding affinity ensemble,
we used Spot instances because they are much cheaper than on-demand.
The downside of a Spot instance is, that it can be terminated at any time,
which can happen if the pool of free Spot instances shrinks over time
and more on-demand capacity is requested in a region.
To minimize instance termination,
we requested a number of instances in each region proportional to the Spot pool size of that region.
We introduced additional flexibility by requesting instances with all possible vCPU counts
and fitting several jobs on them.
A single 96 vCPU \texttt{c5.24xl} instance could then e.g.\ end up running one 48 vCPU job 
plus six eight vCPU jobs at the same time.

\begin{figure}
\caption{{\bf Example of a Docker file for a GPU image.}
From the Docker files multiple Docker container images are compiled (one for each architecture) 
that are loaded from the Amazon ECR by the instances.}
\label{lst:dockergpu}
\begin{lstlisting}[language=Dockerfile]
ARG SRC_IMG=public.ecr.aws/hpc/spack/gromacs/2021.1:cuda-tmpi_linux-amzn2-skylake_avx512-2021-04-29
FROM ${SRC_IMG}

ENV NVIDIA_DRIVER_CAPABILITIES=compute

RUN yum install -y python3-pip jq git
RUN pip3 install --target=/opt/view/lib/python3.8/ --no-warn-script-location --upgrade pip 
RUN pip3 install --target=/opt/view/lib/python3.8/ --no-warn-script-location boto boto3 awscli jsonpickle

RUN yum install -y python-pip
RUN pip install awscli
RUN git clone https://github.com/deGrootLab/pmx /usr/local/src/pmx 
RUN yum install -y gcc python-devel
RUN cd /usr/local/src/pmx \
 && pip install .

COPY batch_processor.py /usr/local/bin/batch_processor.py
COPY start.sh /usr/local/bin/start.sh

VOLUME /scratch
WORKDIR /scratch

COPY ti_verlet_l0.mdp /opt/common/
COPY ti_verlet_l1.mdp /opt/common/
COPY ff /opt/common/ff
VOLUME /opt/common

COPY run-gpu.pl /usr/local/bin/run.pl     # make executable before copying in
COPY run.sh /usr/local/bin/run.sh

## Make sure to add -c as spack won't create the environment correctly
ENTRYPOINT ["/bin/bash","--rcfile","/etc/profile","-l", "/usr/local/bin/start.sh"]
\end{lstlisting}
\end{figure}

To better understand the whole setup, let's look at the encircled digits (red) in Fig.~\ref{fig:batchworkflow} 
and follow the lifetime of one of 19,872 jobs from the binding affinity ensemble.
\Circled{1} We submit an example job from a Cloud9 terminal to the HyperBatch entry point.
The job definition file specifies how many vCPUs to allocate, whether to request a GPU,
and which subfolders to use in the S3 input and output buckets for job I/O.
HyperBatch distributes jobs across regions according to 
region weights reflecting the compute capacity of the regions, e.g.\ using the weights
(6, 6, 3, 1, 1, 4) for the regions
(\texttt{us-east-1}, \texttt{us-east-2}, \texttt{us-west-2}, \texttt{ap-southeast-1}, \texttt{ap-northeast-2}, \texttt{eu-west-1})
for GPU jobs. 
Our example job gets distributed to \texttt{eu-west-1} (blue, \Circled{2}),
where it is relayed to a Batch instance \Circled{3} with sufficient free resources (vCPUs, GPUs).
The instance loads the correct Docker image from AWS Elastic Container Registry (ECR)
with preinstalled software for the current architecture \Circled{4},
e.g.\ pmx and the \gromacs executable with the SIMD level matching the CPU,
see Fig.~\ref{lst:dockergpu} for the definition of the Docker file.

\begin{figure}
\caption{Perl script used to launch each of the 19,872 jobs (first part).}
\label{lst:perlone}
\begin{lstlisting}[language=Perl]
#!/usr/bin/env perl
use Cwd;
my $workdir = getcwd;
my $intpr   = $ARGV[0]; # "s3://input-gaff2-water/cdk8/edge_13_14/stateA/eq1/tpr.tpr"
my $outdir  = $ARGV[1]; # "s3://output-gaff2-water/cdk8/edge_13_14/stateA/run1/"
my $topdir  = $ARGV[2]; # "s3://input-gaff2-water/cdk8/edge_13_14"
my $topfile = $ARGV[3]; # "topolA1.top"
my $mdpfile = $ARGV[4]; # "ti_verlet_l0.mdp"
my $ntomp   = $ARGV[5]; # number of threads e.g. "8"
my $nmpi    = 1;        # number of ranks

system("rm -rf ./*");   # Start clean + remove potential leftovers from earlier run
system("aws s3 cp $topdir . --recursive --exclude='state?/*' ");
$ENV{'GMXLIB'} = "/opt/common/ff";

# Maybe this job did already run on another Spot instances that died at some point.
# Retrieve whatever data we stored from that run, and go on from there.
system("aws s3 cp $outdir . --recursive");

if (-s "frame0.gro") {
    print "=== Frame 0 found, starting transitions. ===\n";
} else {
    print "=== Frame 0 not found, equilibration not complete, continuing eq.===\n";

    #############################################################################
    # FIRST PART: run EQUILIBRATION in chunks, occasionally save generated data #
    #############################################################################
    my @count = (1..8);
    for my $iter (@count) {
        if (-s $intpr) {
            print "--- Found $intpr, continuing ... (iteration $iter)\n";
        } else {
            print "--- Copying $intpr ... (iteration $iter)\n";
            system("aws s3 cp $intpr .");
        }
        system("gmx mdrun -ntmpi $nmpi -ntomp $ntomp -s tpr.tpr -npme 0 -quiet -cpi -nsteps 500000");
        system("rm confout.gro"); # we don't need this one 
        
        # save checkpoint + other files generated by the run to S3 in case Spot instance gets interrupted
        system("aws s3 cp md.log        $outdir"); 
        system("aws s3 cp ener.edr      $outdir"); 
        system("aws s3 cp traj.trr      $outdir"); 
        system("aws s3 cp traj_comp.xtc $outdir"); 
        system("aws s3 cp dhdl.xvg      $outdir"); 
        system("aws s3 cp state.cpt     $outdir"); 

        # check number of steps
        my $nsteps = get_step_number( "md.log" );  # (get_step_number not part of this listing)
        if( $nsteps >= 2999999 )
        { last; }
    }
    system("echo 0 | gmx trjconv -quiet -s tpr.tpr -f traj.trr -o frame.gro -sep -ur compact -pbc mol -b 2256");

    # Save the frames in case this Spot instance gets interrupted:
    system("aws s3 cp . $outdir --recursive --exclude='*' --include='frame*.gro' ");
}
\end{lstlisting}
\end{figure}

\begin{figure}
\caption{Perl script used to launch each of the 19,872 jobs (cont'd).}
\label{lst:perltwo}
\begin{lstlisting}[language=Perl, firstnumber=67]
#############################################################################
# SECOND PART: run the 80 transitions                                       #
#############################################################################
system("mkdir $workdir/morphes");  # create folder to run the transitions in
for(my $i=0; $i<=79; $i++)         # loop over each transition (= frame)
{
    system("mkdir $workdir/morphes/frame$i");                           # make subfolder for this frame
    if( -e "$workdir/dhdl$i.xvg" )                                      # check whether dhdl already exists
    {                                                                   # if yes,
        system("cp $workdir/dhdl$i.xvg $workdir/morphes/frame$i/.");    # copy dhdl to subfolder
        next;                                                           # and proceed to next frame
    }    
    system("mv $workdir/frame$i.gro $workdir/morphes/frame$i/frame$i.gro"); # mv input to subfolder
    chdir("$workdir/morphes/frame$i");                                      # and go there    
    # call grompp and mdrun
    system("gmx grompp -p $workdir/$topfile -c frame$i.gro -f /opt/common/$mdpfile -o ti.tpr -maxwarn 3");
    system("gmx mdrun -ntmpi $nmpi -ntomp $ntomp -s ti.tpr -dhdl dhdl$i.xvg -npme 0");
    system("aws s3 cp dhdl$i.xvg $outdir");                                 # save dhdl to S3
    system("rm *#");                                                        # clean up
}  # done with all 80 dhdl values

# integrate and save work values
chdir("$workdir/morphes");
if( $topfile =~ /A/ )
{  system("pmx analyse --integ_only -fA frame*/dhdl*.xvg -oA work.dat --work_plot none"); }
else
{  system("pmx analyse --integ_only -fB frame*/dhdl*.xvg -oB work.dat --work_plot none"); }
# Copy all results back to correct S3 output container
system("aws s3 cp work.dat $outdir");

exit 0;
\end{lstlisting}
\end{figure}

The actual simulations are handled 
by the Perl script shown in Figs.~\ref{lst:perlone}--\ref{lst:perltwo}.
This script is designed to deal with sudden interrupts that are possible with Spot instances.
Accordingly, output data and checkpoints are saved in regular intervals to S3 storage.
To start a simulation, step \Circled{5} loads the input files from S3
(line 13 in Fig.~\ref{lst:perlone}).
Step \Circled{6} loads potentially existing output data from S3
(line 18 in the listing),
this is the case when the job was running earlier already but was interrupted before it finished.
Depending on whether prior output data is present, the job is either continued or
started from scratch.
Generally, the MD job consists of two parts, 
(i), the production of an equilibrium trajectory (lines 25--56 in the listing), and
(ii), the 80 individual transitions (lines 67--86).
Part (i) is executed in eight chunks (lines 28--29)
so that upon instance termination only a small part needs to be recomputed,
as \Circled{7} each chunk's data is transferred to S3.
If an instance terminates during one of the 80 transitions,
the job is continued from the start of that transition,
as a completed transition \Circled{7} is immediately saved to S3. 
At last, pmx integrates and saves the work values that are later used for free energy estimation (lines 88--95).
Instance termination \Circled{8} at any time will trigger a Lambda function that resubmits
the job again to HyperBatch.

The current state of each job can be checked in a DynamoDB table (Fig.~\ref{fig:batchworkflow} right).
Additional configuration using Amazon Elasticsearch allows
to globally monitor the whole simulation ensemble in a Kibana\cite{gupta2015kibana} dashboard
that shows the total number of running instances, the instance types by region, and more.


\section{Results and Discussion}
We present our results in four parts.
(i) Performance, scaling, and cost efficiency in terms of performance to price (P/P) ratios for the standard MD input systems such as MEM and RIB
on CPU and GPU instances.
(ii) A cost comparison of cloud computing versus buying and operating an own cluster.
(iii) As a prerequisite for the binding affinity study, 
the results of the free energy benchmarks (SHP-2, c-Met, and HIF-2$\alpha$) on various instance types,
including the resulting performance to price ratios.
(iv) The performance and the costs of the binding affinity studies on global cloud resources.

\subsection{Which instances are optimal for \gromacs?}
Tables~\ref{tab:numbers2020}--\ref{tab:g4dn.16xl_scaling2020} show the benchmark results for various instance types.
For CPU instances, Tab.~\ref{tab:numbers2020} lists MEM and RIB performances in gray and blue
colors, and the resulting P/P ratios from greens over yellows to reds, 
corresponding to high, medium and low cost-efficiency.
Tab.~\ref{tab:gpu_instances} shows the same for instances with up to 8 GPUs.
As the mapping of colors to values depends on the smallest and largest observed values 
it differs between MEM and RIB but is the same across all tables:
In result, greens will always refer to good choices in terms of P/P ratio.
For several of the instances, various ranks \mal threads decompositions are listed;
``PME ranks'' indicates if and how many MPI ranks were set aside for PME.

\subsubsection{Performance on individual instances with CPUs}
The highest performances were measured on \texttt{c6i.32xl} and \texttt{m6i.32xl} instances, 
which is expected as with 128 vCPUs they offer the largest number of cores (see also Tab.~\ref{tab:instances}).
Performance-wise, they are followed by 96 vCPU \texttt{c5.24xl} and \texttt{m5n.24xl} instances.
In terms of cost-efficiency, the 72 vCPU \texttt{c5} instances as well as
the 64 vCPU ARM-based \texttt{c6g}'s are a good choice for \gromacs, 
whereas the \texttt{c6i}'s with 128 vCPUs score high for the large MD system.

\begin{table}
\caption{{\bf \gromacs 2020 performance on selected CPU instances.}
\nsday values list MEM and RIB performances,
(ns/\$) columns performance to price.
Values are color-coded for a quick visual orientation:
Grays for low performances, blue towards higher values.
For the performance to price ratios, reds indicate sub-average ratios,
yellows average, and greens above-average ratios.
}
\vspace{-5mm}
\label{tab:numbers2020}
\begin{center}
\scriptsize
\STautoround*{2}
\begin{spreadtab}{{tabular}{ll
ccMLON}}
\toprule
@ processor(s) and        & @price   & @ranks \mal      & @PME   & @{MEM}      & @{RIB}      & @{MEM}                           & @{RIB}                             \\ 
@ \texttt{instance type}  & @(\$/h)  & @\spc threads    & @ranks & @{(\nsday)} & @{(\nsday)} & @{(ns/\$)}                       & @{(ns/\$/10)}                      \\ \midrule
@\texttt{c5.24xl}         &  4.08    & @96 \mal  1      & @  -   &   97.3235   &    6.9465   & \STcopy{v99}{[-2,0]/(24*[-5,0])} & \STcopy{v99}{10*[-2,0]/(24*[-6,0])} \\
@\spc Intel 8275CL        &  4.08    & @48 \mal  2      & @  -   &  105.3735   &    6.3025   &                                  &                                    \\
@\spc 2 x 24 cores        &  4.08    & @48 \mal  1      & @  -   &   96.961    &    6.3435   &                                  &                                    \\
@\spc 96 vCPUs            &  4.08    & @32 \mal  3      & @  -   &   93.8795   &    6.136    &                                  &                                    \\
                          &  4.08    & @24 \mal  4      & @  -   &   96.6770   &    5.9715   &                                  &                                    \\
                          &  4.08    & @16 \mal  6      & @  -   &   91.1270   &    5.726    &                                  &                                    \\
                          &  4.08    & @12 \mal  8      & @  -   &   87.021    &    5.5755   &                                  &                                    \\
                          &  4.08    & @ 8 \mal 12      & @  -   &   81.370    &    5.618    &                                  &                                    \\
                          &  4.08    & @96 \mal  1      & @ 32   &   99.8895   &    6.4925   &                                  &                                    \\
                          &  4.08    & @48 \mal  2      & @ 16   &   91.448    &    6.2555   &                                  &                                    \\
                          &  4.08    & @48 \mal  1      & @ 16   &   92.3805   &    5.9115   &                                  &                                    \\
                          &  4.08    & @32 \mal  3      & @ 12   &   83.318    &    6.3585   &                                  &                                    \\
                          &  4.08    & @24 \mal  2      & @  8   &   89.27     &    5.845    &                                  &                                    \\ \midrule
@\texttt{c5.18xl}         &  3.06    & @36 \mal  1      & @  -   &   81.733    &    4.4545   &                                  &                                    \\
@\spc Intel 8124M         &  3.06    & @72 \mal  1      & @  -   &   86.384    &    4.7995   &                                  &                                    \\
@\spc 2 x 18 cores        &  3.06    & @36 \mal  2      & @  -   &   89.388    &    4.594    &                                  &                                    \\
@\spc 72 vCPUs            &  3.06    & @24 \mal  3      & @  -   &   79.209    &    4.4295   &                                  &                                    \\
                          &  3.06    & @18 \mal  4      & @  -   &   79.346    &    4.349    &                                  &                                    \\
                          &  3.06    & @12 \mal  6      & @  -   &   74.4755   &    4.198    &                                  &                                    \\
                          &  3.06    & @36 \mal  2      & @ 12   &   79.3355   &    5.0005   &                                  &                                    \\
                          &  3.06    & @72 \mal  1      & @ 24   &   84.1015   &    5.094    &                                  &                                    \\ \midrule
@\texttt{m5zn.12xl}       &  3.9641  & @24 \mal  2      & @  8   &   69.700    &    4.620    &                                  &                                    \\
@\texttt{m5n.24xl}        &  5.712   & @96 \mal  1      & @ 32   &  102.147    &    5.895    &                                  &                                    \\ \midrule
@\texttt{c6i.32xl}        &  5.44    & @128\mal  1      & @ 44   &  118.0965   &   10.0415   &                                  &                                    \\
@\texttt{m6i.32xl}        &  6.144   & @40 \mal  1      & @ 24   &  121.866    &   10.076    &                                  &                                    \\ \midrule
@\texttt{c5a.24xl}        &  3.696   & @48 \mal  1      & @  -   &   67.151    &    4.218    &                                  &                                    \\
@\spc AMD EPYC 7R32       &  3.696   & @96 \mal  1      & @  -   &   69.121    &    4.239    &                                  &                                    \\
@\spc 48 cores            &  3.696   & @48 \mal  2      & @  -   &   71.116    &    4.034    &                                  &                                    \\
@\spc 96 vCPUs            &  3.696   & @24 \mal  3      & @  -   &   54.175    &    3.980    &                                  &                                    \\
                          &  3.696   & @24 \mal  4      & @  -   &   64.981    &    4.006    &                                  &                                    \\
                          &  3.696   & @16 \mal  6      & @  -   &   53.326    &    3.786    &                                  &                                    \\
                          &  3.696   & @48 \mal  2      & @ 12   &   67.187    &    4.734    &                                  &                                    \\
                          &  3.696   & @96 \mal  1      & @ 16   &   75.0465   &    5.0155   &                                  &                                    \\
                          &  3.696   & @96 \mal  1      & @ 24   &   76.882    &    4.950    &                                  &                                    \\ \midrule
@\texttt{c6g.16xl}        &  2.176   & @ 1 \mal 64      & @  -   &   54.6525   &    3.5465   &                                  &                                    \\
@\spc ARM Graviton2       &  2.176   & @64 \mal  1      & @  -   &   59.595    &    3.532    &                                  &                                    \\
@\spc 64 cores            &  2.176   & @64 \mal  1      & @ 10   &   50.6255   &    3.3105   &                                  &                                    \\
@\spc 64 vCPUs            &  2.176   & @64 \mal  1      & @ 14   &   62.023    &    3.6585   &                                  &                                    \\
                          &  2.176   & @64 \mal  1      & @ 16   &   51.77     &    3.731    &                                  &                                    \\
                          &  2.176   & @32 \mal  2      & @  -   &   56.2035   &    3.319    &                                  &                                    \\
\bottomrule
\end{spreadtab}
\end{center}
\end{table}

\subsubsection{Performance on single instances with GPUs}
From the GPU instances (Tab.~\ref{tab:gpu_instances}) the ones with a single V100 or T4 GPU 
reach about the performance of the \texttt{c5.24xl} CPU instance, 
albeit with a significantly (1.2--1.8\mal) better cost-efficiency.
In fact, the single-GPU \texttt{g4dn}'s exhibit the best cost-efficiency of all instances
for the MEM and RIB benchmarks.

Perhaps unsurprisingly, the highest single-instance performances of this whole study
have been measured on instances with four and eight GPUs.
With the exception of the (comparatively cheap) quadruple-GPU \texttt{g4dn.12xl} instances, however,
the P/P ratio plunges when distributing a simulation across many GPUs on an instance.
In those cases, \gromacs uses both domain decomposition via MPI ranks as well as OpenMP parallelization,
with added overheads of both approaches.
Additionally, as the PME long range contribution can not (yet) be distributed to multiple GPUs,
it is offloaded to a single GPU, while the other GPUs share the remaining calculations of the nonbonded interactions.
All imbalances in computational load between the GPUs or between the CPU and GPU part
translate into a loss in efficiency and thus in a reduced cost-efficiency.

For single-GPU simulations \gromacs has a performance sweet spot.
Here, domain decomposition is usually not needed nor invoked, and all nonbonded interactions 
including PME can be offloaded to a single GPU, leading to considerably less
imbalance than in the multi-GPU scenario.

To use instances with $N$ GPUs more efficiently, one can run $N$ simulations simultaneously
on them via \gromacs' built-in \texttt{-multidir} functionality,
thus essentially gaining the efficiency of the single-GPU case.
This is demonstrated in Tab.~\ref{tab:gpu_instances} for the \texttt{p4d.24xl} and the \texttt{g4dn.12xl} instances.
The \texttt{p4d.24xl} line in the table shows the results for parallelizing a single simulation across the whole instance,
whereas \texttt{p4d.24xl/8} shows what happens when eight simulations run concurrently.
Here, the produced amount of trajectory and thus also the cost-efficiency, is about four times as high.
For the \texttt{g4dn.12xl/4} vs.\ \texttt{g4dn.12xl} instance, 
running four concurrent simulations instead of one simulation translates into about a factor of two higher cost-efficiency.

\begin{table}
\caption{{\bf \gromacs 2020 performance on individual instances with GPUs.}
As Tab.~\ref{tab:numbers2020}, but on instances with up to eight GPUs.
PME long-range interactions were offloaded to a GPU in all cases, except \stern, where they were evaluated on the CPU.
}
\label{tab:gpu_instances}
\begin{center}
\scriptsize
\STautoround*{2}
\begin{spreadtab}{{tabular}{lclccMLON}}
\toprule
@instance              &@ vCPUs &@ GPU(s)      & @price   &@ ranks \mal                   & @{MEM}      & @{RIB}      & @{MEM}                           & @{RIB}                             \\
@type                  &@       &@             & @(\$/h)  &@ \spc threads                 & @{(\nsday)} & @{(\nsday)} & @{(ns/\$)}                       & @{(ns/\$/10)}                      \\ \midrule
@\texttt{p3.2xl}       &@   8   &@ V100        &  3.06    &@  1 \mal  8                   &  101.287    &    6.4425   & \STcopy{v99}{[-2,0]/(24*[-4,0])} & \STcopy{v99}{10*[-2,0]/(24*[-5,0])}\\
@\texttt{p3.8xl}       &@  32   &@ V100\mal{4} & 12.24    &@  4 \mal  8                   &  142.5475   &    6.824    &                                  &                                    \\
@\texttt{p3.16xl}      &@  64   &@ V100\mal{8} & 24.48    &@  8 \mal  8                   &  202.636    &   13.1395   &                                  &                                    \\
@\texttt{p3.24xl}      &@  96   &@ V100\mal{8} & 31.218   &@  8 \mal 12                   &  227.323    &   16.9815   &                                  &                                    \\ \midrule
@\texttt{p4d.24xl}/8   &@  12   &@ A100        & 32.7726/8&@  1 \mal 12                   &  130.454375 &    7.241375 &                                  &                                    \\
@\texttt{p4d.24xl}     &@  96   &@ A100\mal{8} & 32.7726  &@  8 \mal 12                   &  227.2065   &   15.0935   &                                  &                                    \\ \midrule
@\texttt{g3s.xl}       &@   4   &@ M60         &  0.75    &@  1 \mal  4                   &   37.4390   &    2.089    &                                  &                                    \\
@\texttt{g3.4xl}       &@  16   &@ M60         &  1.14    &@  1 \mal 16                   &   51.1510   &    2.6355   &                                  &                                    \\ \midrule
@\texttt{g4dn.xl}      &@   4   &@ T4          &  0.526   &@  1 \mal  4                   &   57.8825   &    3.173    &                                  &                                    \\
@\texttt{g4dn.2xl}     &@   8   &@ T4          &  0.752   &@  1 \mal  8                   &   76.4985   &    4.0275   &                                  &                                    \\
@\texttt{g4dn.12xl}/4  &@  12   &@ T4          &  3.912/4 &@  1 \mal 12                   &  321.835/4  &   16.377/4  &                                  &                                    \\
@\texttt{g4dn.4xl}     &@  16   &@ T4          &  1.204   &@  1 \mal 16                   &   91.985    &    4.6315   &                                  &                                    \\
@\texttt{g4dn.8xl}     &@  32   &@ T4          &  2.176   &@  1 \mal 32\stern             &  100.085    &    6.339    &                                  &                                    \\
@\texttt{g4dn.16xl}    &@  64   &@ T4          &  4.352   &@  1 \mal 32\stern / 16 \mal 4\stern & 109.5595 & 8.4725   &                                  &                                    \\
@\texttt{g4dn.12xl}    &@  48   &@ T4\mal{4}   &  3.912   &@  4 \mal 12                   &  140.607    &    9.0415   &                                  &                                    \\
\bottomrule
\end{spreadtab}
\end{center}
\end{table}

\subsubsection{Scaling across multiple instances}
For selected instance types, we also determined how much performance can be gained on multiple instances.
For this we have selected instance types that 
(i) exhibit above average P/P ratios for the single-instance benchmarks, and
(ii) have a network speed of at least 50 Gigabit/s.

\begin{table}
\caption{{\bf Scaling across multiple CPU instances.}
\gromacs 2020 performances for MEM and RIB over multiple \texttt{c5n.18xl} instances.
Third column lists the optimal decomposition into MPI ranks and OpenMP threads,
forth column lists the optimal number of separate PME ranks,
left entry for MEM, right entry for RIB if they differ.
}
\label{tab:c5n_scaling2020}
\begin{center}
\STautoround*{3}
\begin{spreadtab}{{tabular}{
S[table-format=2.0,round-mode=places,round-precision=0]
S[table-format=2.0,round-mode=places,round-precision=0]
c
c
S[table-format=4.1,round-mode=places,round-precision=1]
S[table-format=3.2,round-mode=places,round-precision=2]
S[table-format=4.2,round-mode=places,round-precision=2]
S[table-format=3.2,round-mode=places,round-precision=2]
}}
\toprule
@{instan-} &@ total &@ ranks \mal               &@{PME}     & @{MEM}      &@{$E_\text{MEM}$}                     & @{RIB}      &@{$E_\text{RIB}$}                    \\
@{ces}     &@ vCPUs &@ \spc threads             &@{ranks}   & @{(\nsday)} &@                                     & @{(\nsday)} &@                                    \\ \midrule
    1      &    72  &@  36 \mal 2  / 72 \mal 1  &@  0 / 24  &   89.388    & \STcopy{v99}{[-1,0]/([-5,0]*89.388)} &     5.094   & \STcopy{v99}{[-1,0]/([-7,0]*5.094)} \\
    2      &   144  &@  72 \mal 2               &@ 24 / 0   &  105.52     &                                      &     9.6605  &                                     \\
    4      &   288  &@  48 \mal 6 / 144 \mal 2  &@ 16 / 0   &  116.923    &                                      &    17.5395  &                                     \\
    8      &   576  &@  288 \mal 2              &@ 96       &  168.8755   &                                      &    35.831   &                                     \\
   16      &  1152  &@  192 \mal 3 / 576 \mal 2 &@ 64 / 192 &  126.172    &                                      &    55.5490  &                                     \\
   32      &  2304  &@  384 \mal 3 / 576 \mal 2 &@126 / 192 &  109.353    &                                      &    71.4125  &                                     \\
\bottomrule
\end{spreadtab}
\end{center}
\end{table}

Tables~\ref{tab:c5n_scaling2020}, \ref{tab:g4dn.8xl_scaling2020}, and \ref{tab:g4dn.16xl_scaling2020} 
summarize the results for scaling across 1--32 CPU and GPU instances.
For the 81 k atom MEM system, the maximal performance is reached on 8 \texttt{c5n.18xl} instances,
however at a parallel efficiency of less than 25\%, whereas
for the \texttt{g4dn}'s, the highest performance is recorded on individual instances.

In contrast, the large RIB system shows a decent scaling behavior.
On \texttt{c5n.18xl}, the single-instance performance of 5 \nsday can be increased to about
36 \nsday at a parallel efficiency of 88\% on eight instances. 
On 32 instances, with 71 \nsday, the single-instance performance is increased 14-fold.
Whereas the RIB system continues to scale beyond eight \texttt{c5n.18xl} instances,
the \texttt{g4dn}'s never reach 30 \nsday. 

The difference in scaling efficiency between CPU and GPU instances
is mainly determined by the network speed for the inter-node communication.
As the \texttt{c5n.18xl} instances have a much better interconnect than \texttt{g4dn} (see Tab.~\ref{tab:instances}),
the scaling is more efficient for the CPU nodes.

The \texttt{c5n.18xl} instances, however, never reach the scaling performance of an on-premises dedicated HPC cluster.
There, as shown in  Fig.~7 of Ref.~\cite{pall2014}, the same benchmark systems exhibit
peak performances of 303 \nsday for MEM and 204 \nsday for RIB.

\begin{table}
\caption{{\bf Scaling across multiple GPU instances.}
As Tab.~\ref{tab:c5n_scaling2020}, but for \texttt{g4dn.8xl} instances with hyperthreading off.}
\label{tab:g4dn.8xl_scaling2020}
\begin{center}
\STautoround*{3}
\begin{spreadtab}{{tabular}{
S[table-format=2.0,round-mode=places,round-precision=0]
S[table-format=2.0,round-mode=places,round-precision=0]
c
S[table-format=4.1,round-mode=places,round-precision=1]
S[table-format=3.2,round-mode=places,round-precision=2]
S[table-format=4.2,round-mode=places,round-precision=2]
S[table-format=3.2,round-mode=places,round-precision=2]
}}
\toprule
@{instan-} &@{total}&@ ranks \mal                & @{MEM}      &@{$E_\text{MEM}$}                     & @{RIB}      &@{$E_\text{RIB}$}                    \\
@{ces}     &@{cores}&@ \spc threads              & @{(\nsday)} &@                                     & @{(\nsday)} &@                                    \\ \midrule
    1      &    16  &@   1 \mal 16 / 16 \mal 1   &   95.276    & \STcopy{v99}{[-1,0]/([-4,0]*95.276)} &   5.1465    & \STcopy{v99}{[-1,0]/([-6,0]*5.1465)}\\
    2      &    32  &@   4 \mal  8               &   64.964    &                                      &   8.485     &                                     \\
    4      &    64  &@   8 \mal  8 / 32 \mal 2   &   73.0665   &                                      &  15.803     &                                     \\
    8      &   128  &@  32 \mal  4 / 64 \mal 2   &   63.7325   &                                      &  21.2535    &                                     \\
   16      &   256  &@  32 \mal  8               & @           &  @                                   &  25.859     &                                     \\
   32      &   512  &@  32 \mal 16               & @           &  @                                   &  22.781     &                                     \\
\bottomrule
\end{spreadtab}
\end{center}
\end{table}

\begin{table}
\caption{{\bf Scaling across multiple GPU instances.}
As Tab.~\ref{tab:c5n_scaling2020}, but for \texttt{g4dn.16xl} instances with hyperthreading off.}
\label{tab:g4dn.16xl_scaling2020}
\begin{center}
\STautoround*{3}
\begin{spreadtab}{{tabular}{
S[table-format=2.0,round-mode=places,round-precision=0]
S[table-format=2.0,round-mode=places,round-precision=0]
c
S[table-format=4.1,round-mode=places,round-precision=1]
S[table-format=3.2,round-mode=places,round-precision=2]
S[table-format=4.2,round-mode=places,round-precision=2]
S[table-format=3.2,round-mode=places,round-precision=2]
}}
\toprule
@{instan-} &@{total}&@ ranks \mal                & @{MEM}      &@{$E_\text{MEM}$}                     & @{RIB}      &@{$E_\text{RIB}$}                    \\
@{ces}     &@{cores}&@ \spc threads              & @{(\nsday)} &@                                     & @{(\nsday)} &@                                    \\ \midrule
    1      &    32  &@   1 \mal 32 / 8 \mal  4   &   98.0550   & \STcopy{v99}{[-1,0]/([-4,0]*98.055)} &    7.481    & \STcopy{v99}{[-1,0]/([-6,0]*7.481)} \\
    2      &    64  &@   8 \mal  8 / 32 \mal 2   &   75.974    &                                      &  13.2685    &                                     \\
    4      &   128  &@   8 \mal 16 / 32 \mal 4   &   73.187    &                                      &  19.4985    &                                     \\
    8      &   256  &@   32 \mal 8               & @           & @                                    &  24.3895    &                                     \\
   16      &   512  &@   64 \mal 8               & @           & @                                    &  28.3745    &                                     \\
   32      &  1024  &@   32 \mal 32              & @           & @                                    &  21.4735    &                                     \\
\bottomrule
\end{spreadtab}
\end{center}
\end{table}

\begin{figure}
\begin{center}
\includegraphics[width=\textwidth]{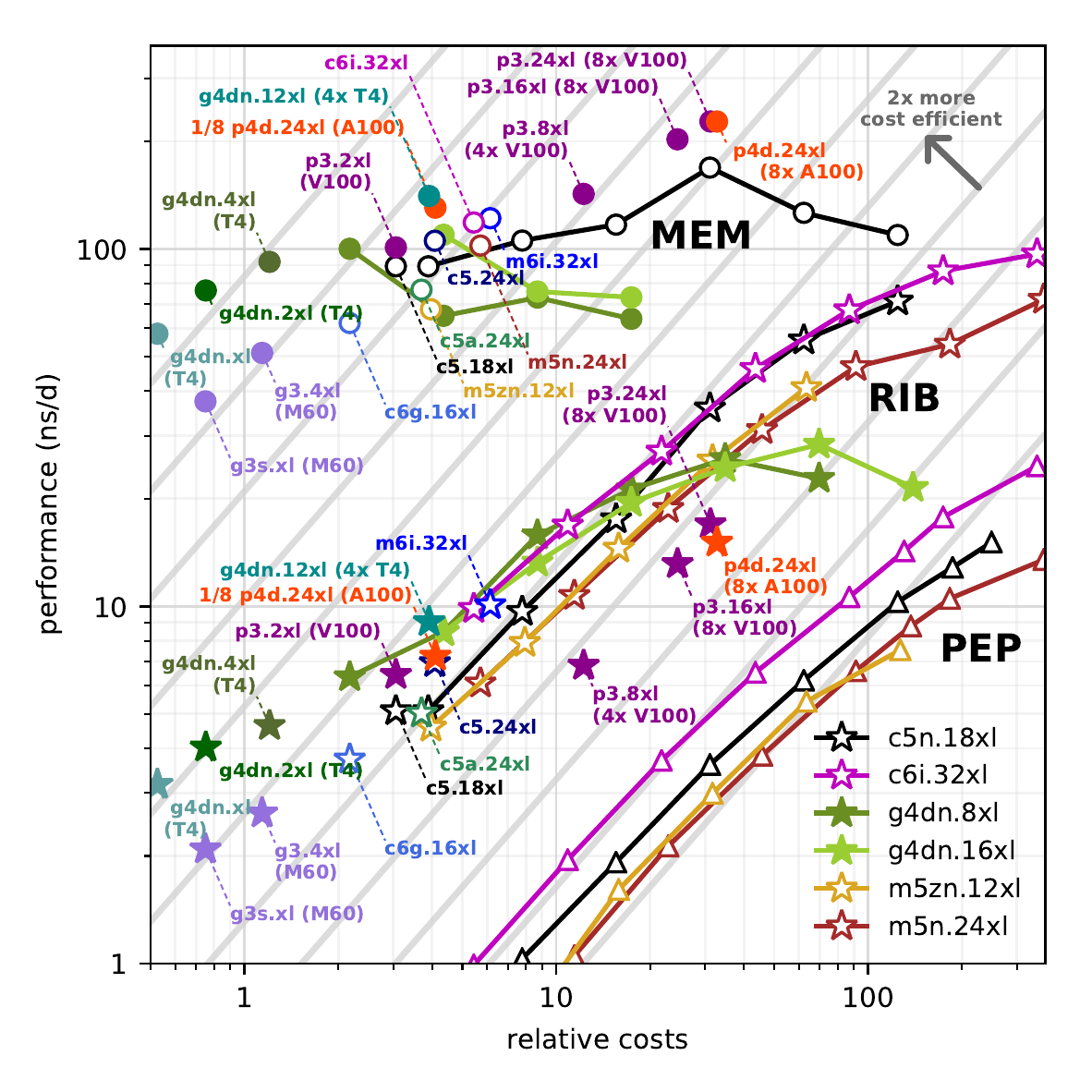}
\caption{{\bf Performance, costs, and cost-efficiency for \gromacs simulations on various AWS instance types.}
\gromacs 2020 performance as a function of the on-demand instance costs (\$/h)
for the MEM (circles), RIB (stars), and PEP (triangles) benchmark on CPU (open symbols) and GPU instances (filled symbols).
Separate symbols indicate single-instances, 
connected symbols show the parallel scaling across multiple instances.}
\label{fig:scalingAWS}
\end{center}
\end{figure}

Fig.~\ref{fig:scalingAWS} summarizes all benchmark results and interrelates them to uncover
which instances are optimal in terms of both performance and cost-efficiency.
The symbols show benchmark performances (at optimal parallelization settings) on various instances
as a function of the on-demand hourly instance costs.
The inclined gray lines are isolines of equal P/P ratio with the most 
cost-efficient configurations towards the upper left.
Moving from one isoline to the neighboring one towards the top left
improves the P/P ratio by a factor of two.
Symbols connected by a line denote the strong scaling behavior across multiple identical instances,
with a single instance at the left end of the curve, followed by 2, 4, 8, and so on, instances. 
A scaling curve that runs parallel to the cost-efficiency isolines
would indicate optimal scaling, i.e.\ a parallel efficiency of $E = 1$.

Fig.~\ref{fig:scalingAWS} allows a series of observations.
(i) In terms of cost-efficiency, the optimal instances for \gromacs are the single-GPU \texttt{g4dn}'s
with 4, 8, and 16 vCPUs (green symbols towards the left),
whose P/P ratio is at least a factor of two higher than most of the other instance types.
(ii) Perhaps unsurprisingly,
the highest MEM and RIB performances on individual instances 
are reached with \texttt{p3} and \texttt{p4d} instances hosting eight GPUs connected via PCI Express (red and purple symbols).
(iii) For larger systems (RIB and PEP), the highest absolute performances are reached
by scaling across multiple \texttt{c6i.32xl} or \texttt{c5n.18xl} instances,
with the \texttt{c6i}'s showing the best cost-efficiency. 
(iv) The performance of small systems like MEM cannot be significantly improved by scaling across many instances.
(v) Choosing one of the many possible instances for an MD project
essentially boils down to pinning down a point along the connecting line 
between best cost-efficiency and highest performance, 
trading off HTC and HPC computing.

Let's follow this special line for the example of the RIB benchmark.
It starts at optimal cost-efficiency with the single-GPU \texttt{g4dn.xl} instances (left, green stars).
For higher performances, one would pick to \texttt{g4dn.2xl} and then \texttt{g4dn.4xl} instances,
however at the cost of losing 20\%--35\% in P/P ratio.
For higher performances (again, at reduced cost-efficiency), the scientist would then 
switch to \texttt{g4dn.16xl}, then \texttt{g4dn.12xl} with 4 GPUs, 
and then continue with scaling over \texttt{c6i} instances (magenta) which exhibit the best P/P ratios
towards growing performances.
There is generally no reason to pick instances within the area below the described line
as here one simply gets lower \gromacs performance for the same price. 
E.g., for the price of a \texttt{g3s} instance (violet, bottom left), 
one would instead get a \texttt{g4dn.2xl} that exhibits two times the RIB performance.

\subsection{Cost comparison: Cloud vs. on-premises cluster}
\label{sec:costComparison}

Whether or not it is more cost-efficient to run simulations on a cloud-based cluster depends of course almost completely on the specific use case,
i.e.\ how big the cluster will be, what software will run on it, 
and whether there are enough jobs at all times to keep the cluster busy
as opposed to bursts of compute demand with idle time in between.
Therefore, no generalisable results or guidance can be provided here. 
We do think however, 
that rough estimates of respective costs and comparison to a typical local compute cluster at a research institution 
will provide useful information and guidelines in particular for new groups in the field who need to setup computer resources.
To this aim, we will estimate and compare the total costs of producing one microsecond of trajectory for the RIB benchmark
with \gromacs.

\begin{figure}
\begin{center}
\includegraphics[width=1.1\textwidth]{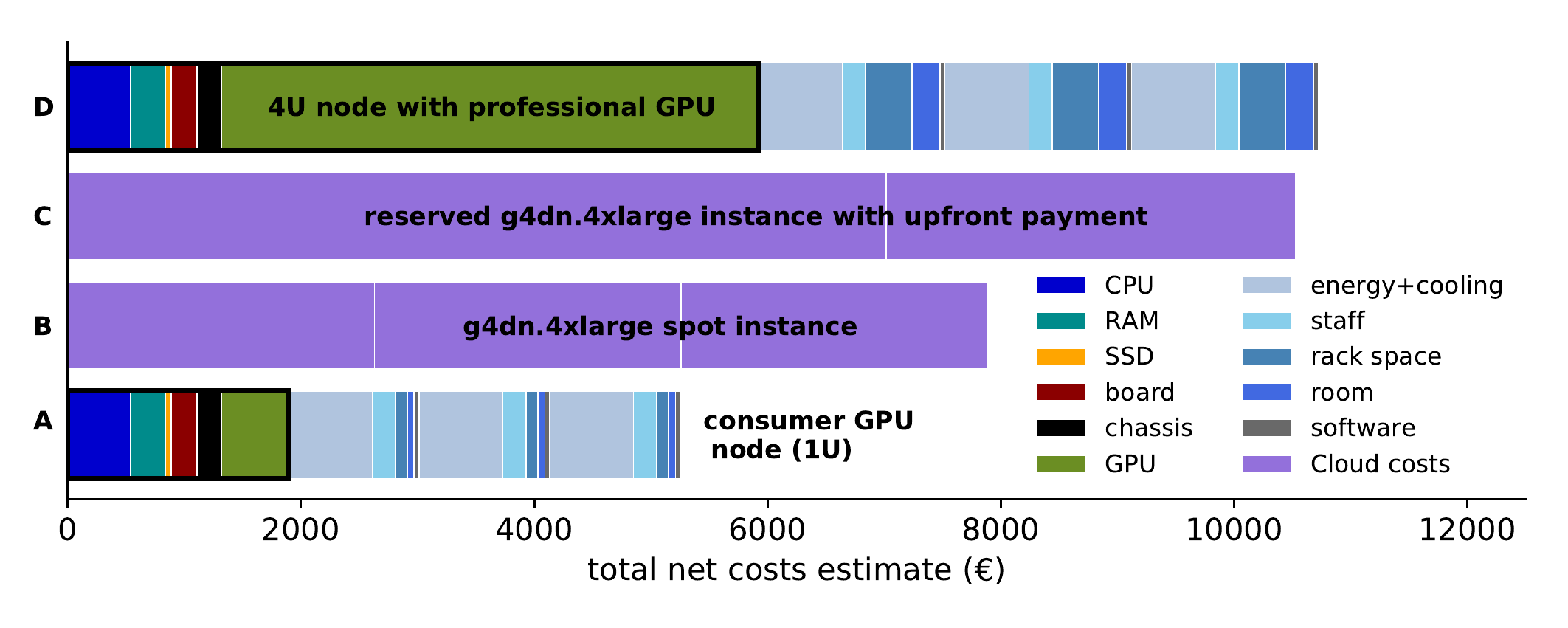}
\caption{\textbf{Costs of a compute node in an owned cluster compared to a cloud instance with similar \gromacs performance over 3 years.}
Violet bars show costs of AWS \texttt{g4dn.4xl} instances (producing 4.63 \nsday of RIB trajectory),
which offer one of the highest performance to price ratios for \gromacs (compare Fig.\ref{fig:scalingAWS}),
in individual blocks of one year.
Bar \textbf{A} shows the fixed costs for buying a consumer GPU node tailored to \gromacs within the thick black line
(broken down into individual hardware components) plus the yearly recurring costs (mainly energy) for three years.
This node (E5-2630v4 CPU plus RTX 2080 GPU) produces 5.9 \nsday of RIB trajectory.\cite{kutznerMoreBang2018}
Bar \textbf{B} shows the average costs using an AWS Spot instance.
Bar \textbf{C} shows the costs when reserving the AWS instance and paying upfront.
Bar \textbf{D} is the same as bar A, but using a 4 U node with a professional GPU (e.g.\ Quadro P6000). 
}
\label{fig:compareTCO}
\end{center}
\end{figure}
The hardware for an own cluster can be aggressively tuned towards cost-efficiency for simulations with \gromacs.
Combining inexpensive processors with consumer GPUs yields the best performance to price ratios.\cite{kutznerMoreBang2018}
For instance, 1~U nodes with an Intel E5-2630v4 processor plus an NVIDIA GeForce RTX 2080 GPU were offered for under 2,000~\Euro net at the time,
including three years of warranty.
Fig.~\ref{fig:compareTCO}A shows a breakdown of the costs into individual contributions for that example.
For nodes similar to those, the RIB trajectory costs can be brought down to approximately 500~\Euro per microsecond
(see Fig.~12 in \cite{kutznerMoreBang2018}).
However, that value is not the total cost of ownership 
as it only reflects the costs for the compute node itself plus energy including cooling,
but no costs for technical staff, room and rack space.

Investment costs for the racks, cooling system, and infrastructure needed to operate the cluster
are estimated to about 500~\Euro per U of rack space over the lifetime of the racks. 
For a lifetime of 5 years that adds 100~\Euro per U per year.
For technical staff to operate, repair, and look after a 500 node cluster,
we assume 100,000~\Euro per year,
which adds 200~\Euro to the operating costs for each node per year.
A suitable room 
(60 -- 100~m$^2$ for about 500~U of hardware with appropriate infrastructure and the possibility to install heavy apparatus)
adds about 30,000~\Euro to the yearly costs (60~\Euro per node), depending on the location.
For cluster management software we assume 40~\Euro per node per year.

Taken together, that adds $100+200+60+40=400$~\Euro for each node per year.
As our exemplary nodes (E5-2630v4 CPU with RTX 2080 GPU) have been benchmarked\cite{kutznerMoreBang2018} to produce
5.9~\nsday of RIB trajectory, a node needs 170 days for a microsecond.
This adds $170/365 * 400 \approx 185$~\Euro to the trajectory production costs.
Including those costs, total RIB trajectory costs can be estimated to be roughly 700~\Euro per microsecond
with optimal hardware.
Bar D of Fig.~\ref{fig:compareTCO} is to illustrate how the costs grow when using the
same hardware as in bar A, but now with a professional GPU (e.g.\ an NVIDIA Quadro P6000) instead of 
a consumer GPU (which leads to considerably higher fixed costs) 
and in a larger chassis that takes 4~U rack space 
(which lead to significantly increased recurring costs for room and rack space).
Thus, densely packed hardware helps to reduce costs.

\texttt{g4dn.4xl} instances offer both a high absolute performance as well as a good 
performance to price ratio for producing RIB trajectory with \gromacs (Fig.~\ref{fig:scalingAWS}),
which would therefore be a good pick for production runs.
As seen in Tab.~\ref{tab:gpu_instances}, one produces 4.63~\nsday for 1.20 dollars (1.00~\Euro) per hour 
on such an instance, i.e.\ 
one microsecond of RIB trajectory would cost about 5,200~\Euro on an on-demand instance.
To reduce costs, one would reserve an instance for one or three years, 
and for maximal savings one can pay upfront. 
In the latter case, a \texttt{g4dn.4xl} would cost about 0.40~\Euro per hour,
translating to about 2,100~\Euro for a microsecond of RIB trajectory.
Running on Spot instances will on average reduce the trajectory cost by 70 percent compared to
the on-demand price, as illustrated in bar B of Fig.~\ref{fig:compareTCO},
resulting in 1,500~\Euro for a microsecond of RIB trajectory.

However, for our trajectory cost comparison we implicitly assume that the department cluster we operate
indeed does produce trajectory 100 percent of the time and there is no down or idle time.
If cluster nodes are producing useful data only 3/4 of the time,
it will increase trajectory costs by 4/3,
whereas for Spot instances you only need to pay when you do need them for productive use time.
For a cluster utilization of 75 \% this would increase the trajectory production costs
of our optimal nodes to about 950 \Euro per microsecond of RIB trajectory.

In summary, with careful selection of cloud resources and payment options, 
there is not much difference in cost today compared to on-premises computing.

\subsection{\gromacs performance for free energy calculations}
Turning on FE perturbations reduces the \gromacs performance, because 
an additional PME grid is evaluated, 
and because interactions involving perturbed atoms run through kernels that are not
as optimized as the standard kernels.
How much the performance differs with and without FE depends on
how big the fraction of perturbed atoms is and
on the parameters chosen for FE.
For those reasons we cannot use the MEM and RIB benchmarks to predict the performances 
of the systems used in our high throughput ligand screening study.
Instead, we carry out new benchmarks for four representative FE systems 
(Tab.~\ref{tab:systems}) chosen from the whole binding affinity ensemble (Tab.~\ref{tab:ba1}).

The performances for these systems, which are a small ligand in water system (from the c-Met dataset)
plus three protein-ligand complexes of different size
(HIF-2$\alpha$, c-Met, and SHP-2) are shown in Tab.~\ref{tab:equil_CPU} for CPU instances
for various decompositions into MPI ranks and OpenMP threads.

The table shows the general trend of small instances exhibiting higher P/P ratios
but there are no pronounced differences between the architectures.
The highest performances are observed on the 96 vCPU Intel instances.

\begin{table}
\caption{\textbf{Free energy benchmarks on CPU instances.}
Performances (\nsday) and performance to price ratios (ns/\$) for \gromacs 2020
on various Intel (\texttt{c5} and \texttt{m5zn}), AMD (\texttt{c5a}), and ARM (\texttt{c6g}) CPU instances.
Color-coding as in Tab.~\ref{tab:numbers2020}.}
\label{tab:equil_CPU}
\begin{center}
\scriptsize
\STautoround*{2}
\begin{spreadtab}{{tabular}{lcccKCXRYQZP}}
\toprule
@                  &@      & @        &@              & @\multicolumn{2}{c}{--- ligand ---}                 & @\multicolumn{6}{c}{--- protein-ligand complex ---}                                                                                                             \\ 
@instance          &@vCPUs & @price   &@ ranks \mal   & @\multicolumn{2}{c}{c-Met}                           & @\multicolumn{2}{c}{HIF-2$\alpha$}                          & @\multicolumn{2}{c}{c-Met}                          & @\multicolumn{2}{c}{SHP-2}                            \\ 
@type              &@      & @(\$/h)  &@ \spc threads & @{(\nsday)}      & @{(ns/\$)}                       & @{(\nsday)}      & @{(ns/\$)}                       & @{(\nsday)}      & @{(ns/\$)}                       & @{(\nsday)}      & @{(ns/\$)}                       \\ \midrule
@\texttt{c5.24xl}  &@  96  &  4.08    &@  2  \mal 48  &                  &                                  &      44.346      & \STcopy{v99}{[-1,0]/(24*[-5,0])} &      27.705      & \STcopy{v99}{[-1,0]/(24*[-7,0])} &     18.379       & \STcopy{v99}{[-1,0]/(24*[-9,0])} \\
@                  &@  96  &  4.08    &@  3  \mal 32  &                  &                                  &      42.535      &                                  &      28.377      &                                  &     19.460       &                                  \\
@                  &@  96  &  4.08    &@  4  \mal 24  &                  &                                  &      55.595      &                                  &      35.252      &                                  &     25.338       &                                  \\
@                  &@  96  &  4.08    &@  6  \mal 16  &                  &                                  &      66.884      &                                  &      37.958      &                                  &     26.872       &                                  \\
@                  &@  96  &  4.08    &@  8  \mal 12  &                  &                                  &      69.454      &                                  &      40.437      &                                  &     28.566       &                                  \\
@                  &@  96  &  4.08    &@ 12  \mal 8   &                  &                                  &      71.678      &                                  &      46.960      &                                  &     30.540       &                                  \\
@                  &@  96  &  4.08    &@ 16  \mal 6   &                  &                                  &      80.647      &                                  &      43.397      &                                  &     34.358       &                                  \\
@                  &@  96  &  4.08    &@ 24  \mal 4   &                  &                                  &      83.176      &                                  &      46.336      &                                  &     32.223       &                                  \\
@                  &@  96  &  4.08    &@ 32  \mal 3   &                  &                                  &      82.789      &                                  &      48.210      &                                  &     36.361       &                                  \\
@                  &@  96  &  4.08    &@ 48  \mal 2   &                  &                                  &      89.649      &                                  &      45.598      &                                  &     34.907       &                                  \\
@                  &@  96  &  4.08    &@ 96  \mal 1   &                  &                                  &      64.936      &                                  &      32.197      &                                  &     20.202       &                                  \\ \cmidrule{1-4}
@\texttt{c5.18xl}  &@  72  &  3.06    &@ 18  \mal 4   &                  &                                  &      65.444      &                                  &      42.165      &                                  &     28.086       &                                  \\
@\texttt{c5.12xl}  &@  48  &  2.04    &@  1  \mal 48  &                  &                                  &      52.355      &                                  &      31.391      &                                  &     21.877       &                                  \\
@\texttt{c5.9xl}   &@  36  &  1.53    &@  1  \mal 36  &                  &                                  &      47.048      &                                  &      26.604      &                                  &     18.055       &                                  \\
@\texttt{c5.4xl}   &@  16  &  0.68    &@  1  \mal 16  &      77.012      & \STcopy{v99}{[-1,0]/(24*[-3,0])} &      27.319      &                                  &      15.089      &                                  &     10.011       &                                  \\
@\texttt{c5.2xl}   &@   8  &  0.34    &@  1  \mal 8   &      52.371      &                                  &      15.236      &                                  &       8.029      &                                  &      5.225       &                                  \\
@\texttt{c5.xl}    &@   4  &  0.17    &@  1  \mal 4   &      30.672      &                                  &@                 &@                                 &@                 &@                                 &@                 &@                                 \\
@\texttt{c5.large} &@   2  &  0.085   &@  1  \mal 2   &      18.106      &                                  &@                 &@                                 &@                 &@                                 &@                 &@                                 \\ \midrule
@\texttt{m5zn.12xl}&@  48  &  3.9641  &@  8  \mal 6   &                  &@                                 &      64.347      &                                  &       36.753     &                                  &     25.501       &                                  \\
@\texttt{m5zn.2xl} &@   8  &  0.6607  &@  1  \mal 8   &       57.663     &                                  &      19.282      &                                  &       10.059     &                                  &      6.568       &                                  \\ \midrule
@\texttt{c5a.24xl} &@  96  &  3.696   &@  48 \mal  2  &                  &@                                 &     71.728       &                                  &       39.964     &                                  &      27.626      &                                  \\
@\texttt{c5a.16xl} &@  64  &  2.464   &@  32 \mal  2  &                  &@                                 &     58.710       &                                  &       32.737     &                                  &      21.054      &                                  \\
@\texttt{c5a.12xl} &@  48  &  1.848   &@  24 \mal  2  &                  &@                                 &     46.065       &                                  &       25.450     &                                  &      16.496      &                                  \\
@\texttt{c5a.8xl}  &@  32  &  1.232   &@  32 \mal  1  &                  &@                                 &     32.966       &                                  &       17.902     &                                  &      11.573      &                                  \\
@\texttt{c5a.4xl}  &@  16  &  0.616   &@   1 \mal 16  &     70.034       &                                  &     18.382       &                                  &        9.988     &                                  &       6.630      &                                  \\
@\texttt{c5a.2xl}  &@   8  &  0.308   &@   1 \mal  8  &     48.891       &                                  &     10.775       &                                  &        5.719     &                                  &       3.698      &                                  \\
@\texttt{c5a.xl}   &@   4  &  0.154   &@   1 \mal  4  &     26.094       &                                  &@                 &@                                 &@                 &@                                 &@                 &@                                 \\
@\texttt{c5a.large}&@   2  &  0.077   &@   1 \mal  2  &     13.414       &                                  &@                 &@                                 &@                 &@                                 &@                 &@                                 \\ \midrule
@\texttt{c6g.16xl} &@  64  &  2.176   &@  32 \mal  2  &                  &@                                 &     58.072       &                                  &       32.846     &                                  &      21.604      &                                  \\
@\texttt{c6g.12xl} &@  48  &  1.632   &@  12 \mal  4  &     151.413      &                                  &     46.263       &                                  &       25.866     &                                  &      17.246      &                                  \\
@\texttt{c6g.8xl}  &@  32  &  1.088   &@   4 \mal  8  &     111.261      &                                  &     32.645       &                                  &       18.745     &                                  &      12.157      &                                  \\
@\texttt{c6g.4xl}  &@  16  &  0.544   &@   1 \mal 16  &      69.803      &                                  &     18.583       &                                  &       10.325     &                                  &      6.579       &                                  \\
@\texttt{c6g.2xl}  &@   8  &  0.272   &@   1 \mal  8  &      42.473      &                                  &     10.043       &                                  &        5.527     &                                  &      3.451       &                                  \\
@\texttt{c6g.xl}   &@   4  &  0.136   &@   1 \mal  4  &      23.319      &                                  &      @           &               @                  &        @         &               @                  &        @         &               @                  \\
\bottomrule
\end{spreadtab}
\end{center}
\end{table}

\begin{figure}[tb]
\begin{center}
\includegraphics[width=0.85\textwidth]{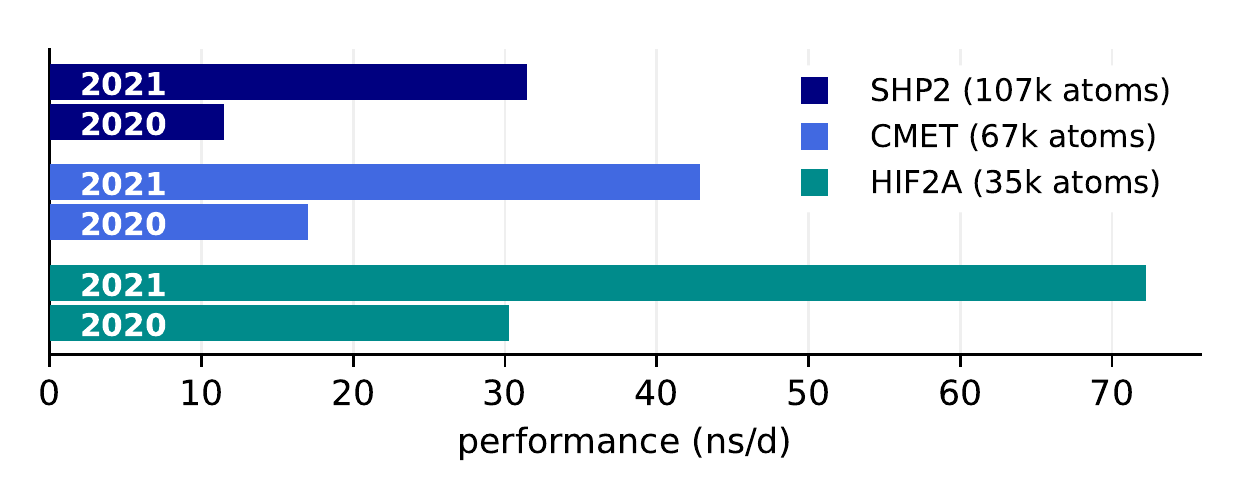}
\caption{\textbf{Performance improvements of \gromacs 2021 for free energy calculations on GPUs.}
For three different MD systems (colors) with free energy perturbation turned on,
the bars compare \gromacs 2021 and 2020 performances
on a \texttt{p3.2xl} instance.
}
\label{fig:performance20vs21}
\end{center}
\end{figure}

Up to version 2020, with perturbed charges it was not possible to offload the PME grid calculations to the GPU.
This has changed from version 2021 on, leading to considerably enhanced performance on GPU instances
of more than a factor of two in our cases (Fig.~\ref{fig:performance20vs21}).
Therefore, on GPU instances, we used \gromacs 2021 for all binding affinity simulations.
The benchmark results for the four representative FE systems on GPU instances are assembled in Tab.~\ref{tab:equil_GPU}. 

\begin{table}[tbp]
\caption{\textbf{Free energy benchmarks on GPU instances.}
As in Tab.~\ref{tab:equil_CPU}, but now for
\gromacs 2021 using one GPU per simulation.
The single-GPU performance on \texttt{p4d.24xl} was derived by running 8 identical benchmarks,
each using one GPU and 1/8th of the hardware threads, in a multi-simulation.
}
\label{tab:equil_GPU}
\begin{center}
\scriptsize
\STautoround*{2}
\begin{spreadtab}{{tabular}{lcccKCXRYQZP}}
\toprule
@                      &@      & @          &@              & @\multicolumn{2}{c}{--- ligand ---}                 & @\multicolumn{6}{c}{--- protein-ligand complex ---}                                                                                                             \\ 
@instance              &@vCPUs & @price     &@ ranks \mal   & @\multicolumn{2}{c}{c-Met}                           & @\multicolumn{2}{c}{HIF-2$\alpha$}                          & @\multicolumn{2}{c}{c-Met}                          & @\multicolumn{2}{c}{SHP-2}                            \\ 
@type                  &@      & @(\$/h)    &@ \spc threads & @{(\nsday)}      & @{(ns/\$)}                       & @{(\nsday)}      & @{(ns/\$)}                       & @{(\nsday)}      & @{(ns/\$)}                       & @{(\nsday)}      & @{(ns/\$)}                       \\ \midrule
@\texttt{p3.2xl}       &@  8   &  3.06      &@  1 \mal  8   &                  &                                  &      72.207      & \STcopy{v99}{[-1,0]/(24*[-5,0])} &      42.830      & \STcopy{v99}{[-1,0]/(24*[-7,0])} &     31.480       & \STcopy{v99}{[-1,0]/(24*[-9,0])} \\
@\texttt{p4d.24xl}/8   &@ 96/8 & 32.7726/8  &@  1 \mal 12   &                  &                                  &      90.451      &                                  &      57.303      &                                  &     41.919       &                                  \\
@\texttt{g3s.xl}       &@   4  &  0.75      &@  1 \mal  4   &     61.415       & \STcopy{v99}{[-1,0]/(24*[-3,0])} &      43.911      &                                  &      22.814      &                                  &     14.211       &                                  \\
@\texttt{g3.4xl}       &@  16  &  1.14      &@  1 \mal 16   &    125.923       &                                  &      60.595      &                                  &      30.554      &                                  &     19.330       &                                  \\ \midrule
@\texttt{g4dn.16xl}    &@  64  &  4.352     &@  1 \mal 64   &     90.100       &                                  &      93.553      &                                  &      64.729      &                                  &     42.260       &                                  \\
@                      &@  64  &  4.352     &@  1 \mal 32   &    119.676       &                                  &     106.986      &                                  &      68.354      &                                  &     43.710       &                                  \\
@\texttt{g4dn.8xl}     &@  32  &  2.176     &@  1 \mal 32   &    130.074       &                                  &     108.721      &                                  &      67.957      &                                  &     43.160       &                                  \\
@                      &@  32  &  2.176     &@  1 \mal 16   &    134.507       &                                  &     114.142      &                                  &      69.215      &                                  &     43.467       &                                  \\
@\texttt{g4dn.4xl}     &@  16  &  1.204     &@  1 \mal 16   &    117.325       &                                  &      96.552      &                                  &      61.866      &                                  &     38.985       &                                  \\
@                      &@  16  &  1.204     &@  1 \mal  8   &    118.946       &                                  &      93.666      &                                  &      56.305      &                                  &     38.397       &                                  \\ 
@\texttt{g4dn.2xl}     &@   8  &  0.752     &@  1 \mal  8   &     98.946       &                                  &      70.448      &                                  &      41.158      &                                  &     29.406       &                                  \\
@                      &@   8  &  0.752     &@  1 \mal  4   &     86.068       &                                  &      69.728      &                                  &      41.422      &                                  &     29.489       &                                  \\ 
@\texttt{g4dn.xl}      &@   4  &  0.526     &@  1 \mal  4   &     58.156       &                                  &      49.401      &                                  &      28.692      &                                  &     19.626       &                                  \\
@                      &@   4  &  0.526     &@  1 \mal  2   &     47.809       &                                  &      42.145      &                                  &      24.075      &                                  &     16.501       &                                  \\
\bottomrule
\end{spreadtab}
\end{center}
\end{table}

Whereas the performances of the 32 and 64 vCPU \texttt{g4dn} instances are comparable to or higher than that of
the best performing CPU instances (i.e.\ \texttt{c6g.12xl} for the ligand in water and \texttt{c5.24xl} for the protein-ligand complexes),
the smaller \texttt{g4dn} instances with $\leq 16$ vCPUs still offer high performances 
but at exceptionally high P/P ratios: about two times higher than on CPU instances.
On the instances with $\geq 32$ vCPUs it is beneficial for performance 
to just use half the number of vCPUs for OpenMP threads,
as the reduction of value over too many threads can deteriorate performance otherwise.

In a nutshell, the highest performances are observed on GPU-accelerated \texttt{g4dn.8xl} instances for the protein-ligand systems,
and on the ARM-based \texttt{c6g.12xl} instances for the small ligand in water systems.
Regarding cost-efficiency, any of the \texttt{c5}, \texttt{c5a}, or \texttt{c6g} instances with $\leq 8$ vCPUs
has a high P/P ratio for the small ligand in water systems, whereas single-GPU
\texttt{g4dn} instances with $\leq 16$ vCPUs are undefeated for the larger protein-ligand systems.

\subsubsection*{Costs and time-to-solution per FE calculation}

\begin{figure}[tbp]
\begin{center}
\includegraphics[width=0.8\textwidth]{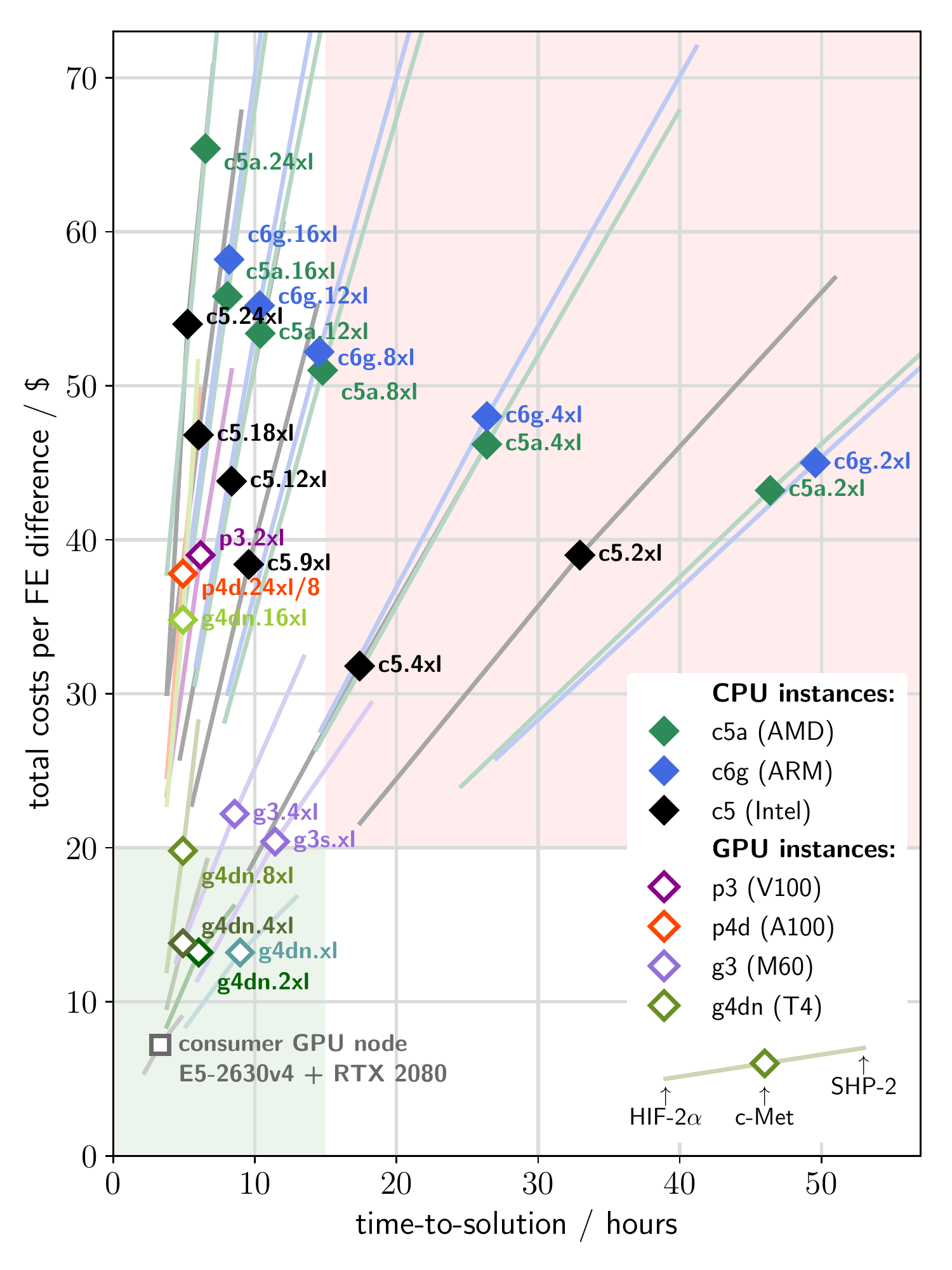}\\
\caption{\textbf{Costs and time needed to compute one FE difference.}
Diamonds show the costs to compute one FE difference (using Spot pricing)
versus the time-to-solution for various instance types (colors) for the c-Met system.
In addition to c-Met, HIF-2$\alpha$ is shown at the lower left end of each colored line,
and SHP-2 at the upper right end.
Gray square shows costs and timings for a consumer GPU node specifically tuned for \gromacs simulations,
as discussed in Sec.~\ref{sec:costComparison} and shown in Fig.~\ref{fig:compareTCO}A.
}
\label{fig:optimalConfig}
\end{center}
\end{figure}

The numbers in Tables \ref{fig:performance20vs21} and \ref{tab:equil_GPU} are for the
equilibration phase of the FE calculation (see Sec.~\ref{sec:methodsFEbench}).
We do not list the benchmark results of the transition phase separately,
but included them in the estimate of the total cost of computing one FE difference,
as described in the methods section.
Fig.~\ref{fig:optimalConfig} shows the time-to-solution and the costs per FE difference that result 
when using different Spot instance types.
With three replicas and two directions,
the total costs for one FE difference is 6\mal the time needed for the protein-ligand part,
plus the (small) costs of the ligand in water.

Spot instance costs are just about a third of the on-demand costs (not shown), 
although Spot prices vary slightly among the regions and change over time.
We therefore used Spot instances for our binding affinity studies,
even though these may be terminated at any time should there be demand for that instance type in the given region.

As can be seen in the Figure, on CPU instances the time-to-solution generally shrinks with the number of vCPUs (as expected) while the costs grow.
Using \texttt{g4dn.xl}, \texttt{g4dn.2xl}, or \texttt{g4dn.4xl} GPU instances, 
any FE calculation is completed within 15 hours for less than 20~\$ for all systems (green quadrant).
Other single-GPU instances like \texttt{g4dn.8xl} and \texttt{g3.4xl} are somewhat less cost-efficient, but
still better than the remaining instance types.
The white quadrant on top of the green quadrant accommodates multiple instance types 
on which a FE value can be computed in less that 15 hours,
albeit at a markedly higher cost than on \texttt{g4dn} instances.

\subsection{High throughput ligand screening in the global cloud}
\subsubsection{Study 1: Focus on time-to-solution}
Our first screening study consisted of 19,872 Batch jobs to compute 1,656 free energy differences 
(200 $\mu$s of trajectory in total) for the ensemble shown in Tab.~\ref{tab:ba1}.
With this study we evaluate the suitability of cloud computing for large scale computational drug design scans
that have been traditionally performed on on-premises clusters where
such a scan would typically take several weeks to complete.

\begin{figure}[tb]
\includegraphics[width=\textwidth]{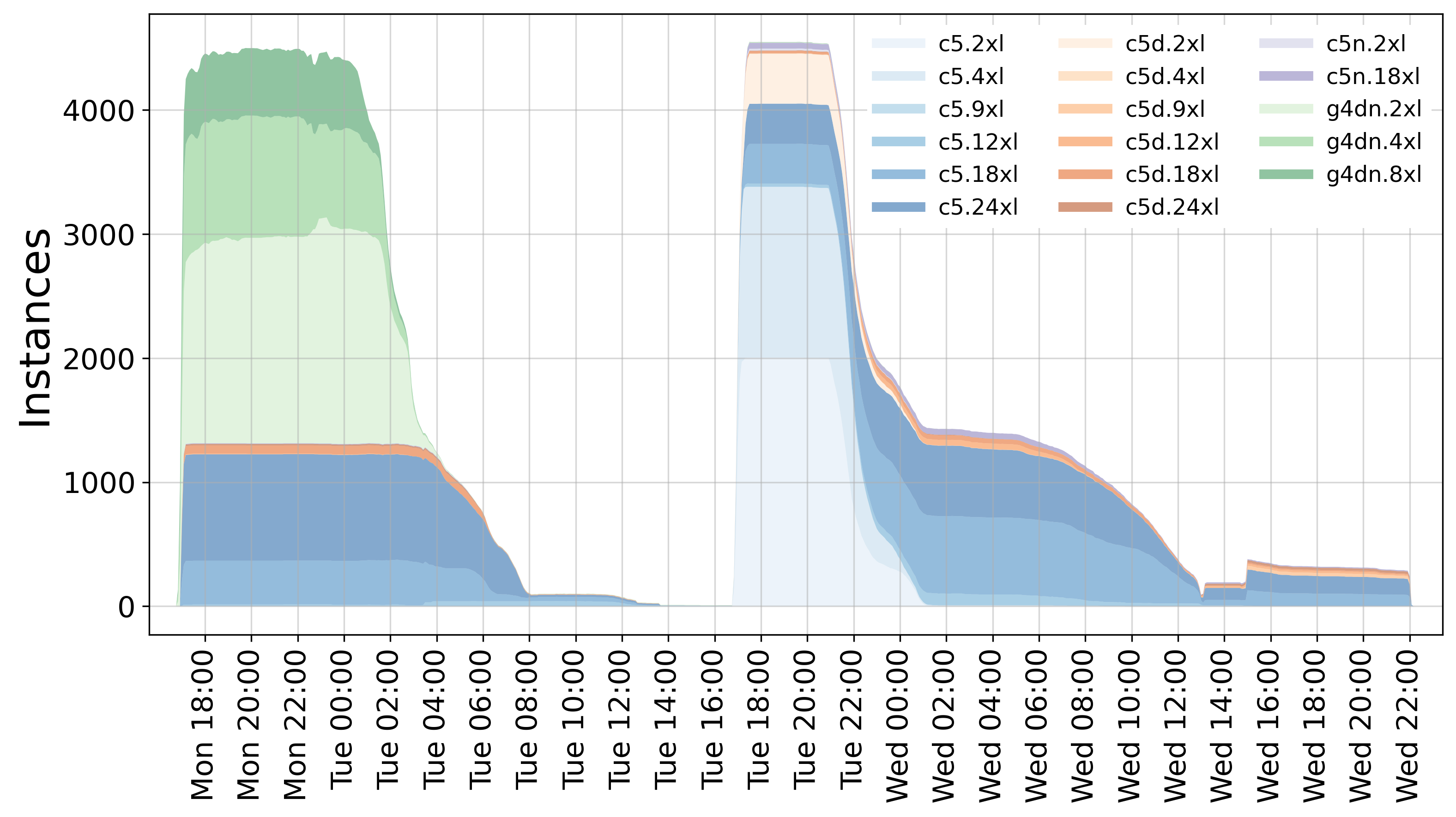}\\
\caption{\textbf{Usage of global compute resources for the first ligand screening study aimed to optimize time-to-solution.}
Colors show the various instances that were in use globally during the three days of the ensemble run.}
\label{fig:instances}
\end{figure}

As we aimed to minimize the time-to-solution, from all available instance types we 
only selected instances that would need no more than nine hours for any job.
The \texttt{g4dn.2xl}, \texttt{g4dn.4xl}, and \texttt{g4dn.8xlarge} meet that criterion
at the lowest costs.
However, relying on just three instance types is risky if one wants to minimize time-to-solution.
\texttt{g4dn} instances are not very abundant in the AWS cloud and if they happen to be in high demand
at the time of our screening study, we might not get many of them.
Therefore, we added other instance types that meet our nine hour criterion, but that 
are almost always available:
\texttt{c5.24xl} and \texttt{c5.18xl} as well as the similar \texttt{c5d.24xl} and \texttt{c5d.18xl}.
We ran the small systems of ligand in water on eight \texttt{c5} vCPUs, 
where they would complete in about five hours at a price of less than 2~\$
and high cost-efficiency (see c-Met ligand column in Tab.~\ref{tab:equil_CPU}).
To draw from a larger pool of instances we allowed for \texttt{c5} instances of various size and
just requested that they offer at least eight vCPUs (see also Fig.~\ref{fig:instances}).
Larger instances accept multiple jobs until they do not have enough free vCPUs left.

\begin{figure}
\begin{center}
\includegraphics[width=0.85\textwidth]{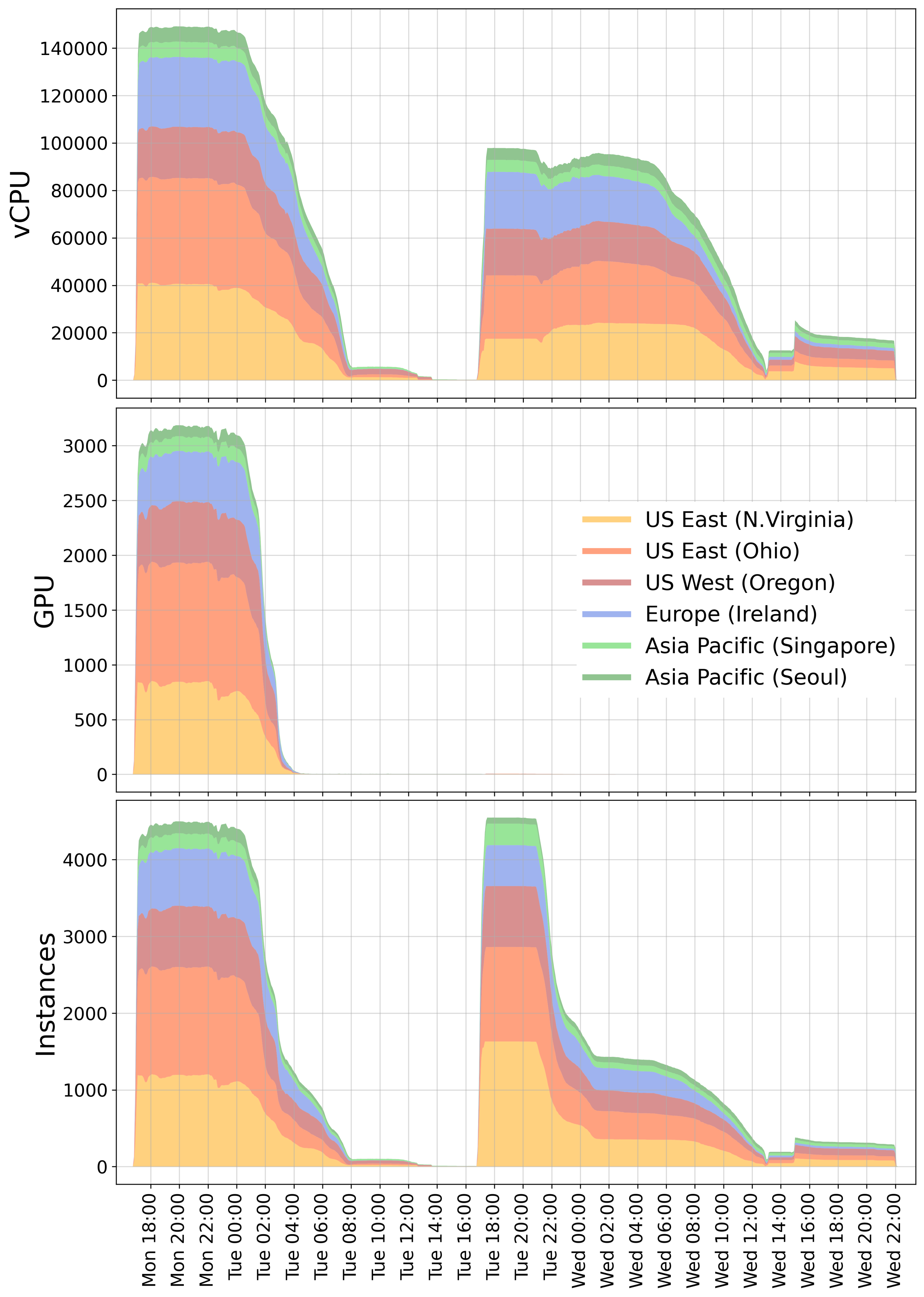}\\
\caption{\textbf{Usage of global compute resources for the first ligand screening study aimed to optimize time-to-solution.}
Compute resources (split into regions) allocated for the ensemble run over time. 
Top: vCPUs, middle: GPU instances, bottom: number of instances.}
\label{fig:bigrun}
\end{center}
\end{figure}

We submitted the first 9,936 jobs (the large protein systems) in a first wave on a Monday at around 5 p.m.,
and the second 9,936 jobs (the small ligand systems) in a second wave 24 hours later.
Fig.~\ref{fig:instances} shows the number of instances that were in use during our first screening study color-coded by instance type.
Fig.~\ref{fig:bigrun} provides further details of the run.
As can be seen from the top and middle panels,
we acquired about 140,000 vCPUs within the first 30 minutes,
and about 3,000 GPUs within the first two hours of the run,
distributed globally over six regions.

Each wave finished in about one day,
and we speculate that also the whole 19,872 jobs would have finished within 24 hours if submitted simultaneously.
As GPU instance availability is essentially independent of the CPU instance availability,
the GPU jobs from the first wave (greens in Fig.~\ref{fig:instances}) can completely overlap with the
CPU jobs of the second wave.
At the same time, after the peak of the second wave (Tue 17 h -- 23 h), there
should be more than enough \texttt{c5} Spot capacity to accommodate the CPU jobs of the first wave.

Unfortunately there was a glitch in the initial version of our setup that prevented finished instances to terminate properly.
For that reason, the actual costs of the first run summed up to 56 \$ per FE difference,
although, when counting productive time only, they reduce to 40 \$ per FE difference.
This is in the expected range (see Fig.~\ref{fig:optimalConfig}), given the mix of instances that were in use.
The overall costs almost entirely resulted from the hourly charges of the EC2 compute instances,
whereas data transfer to and from the S3 buckets accounted for less than 0.5 \% of the whole costs.

In addition to the performance and price evaluation, we have also validated correctness of the calculations 
against the previous computations.~\cite{gapsys2022raven}
We used Student's t-test to compare free energy values calculated in the current work to those reported previously
ensuring that the results showed no statistically significant differences.

\subsubsection{Study 2: Focus on cost-efficiency}
Our second screening study aimed to further reduce the price tag by incorporating only the most cost-efficient instance types for the run.
The second study used a slightly different and smaller data set (see Tab.~\ref{tab:ba2}) that required about 6,984 jobs to be run
for 582 FE differences, or 70 $\mu$s of trajectory in total.

\begin{figure}[tbp]
\includegraphics[width=.9\textwidth]{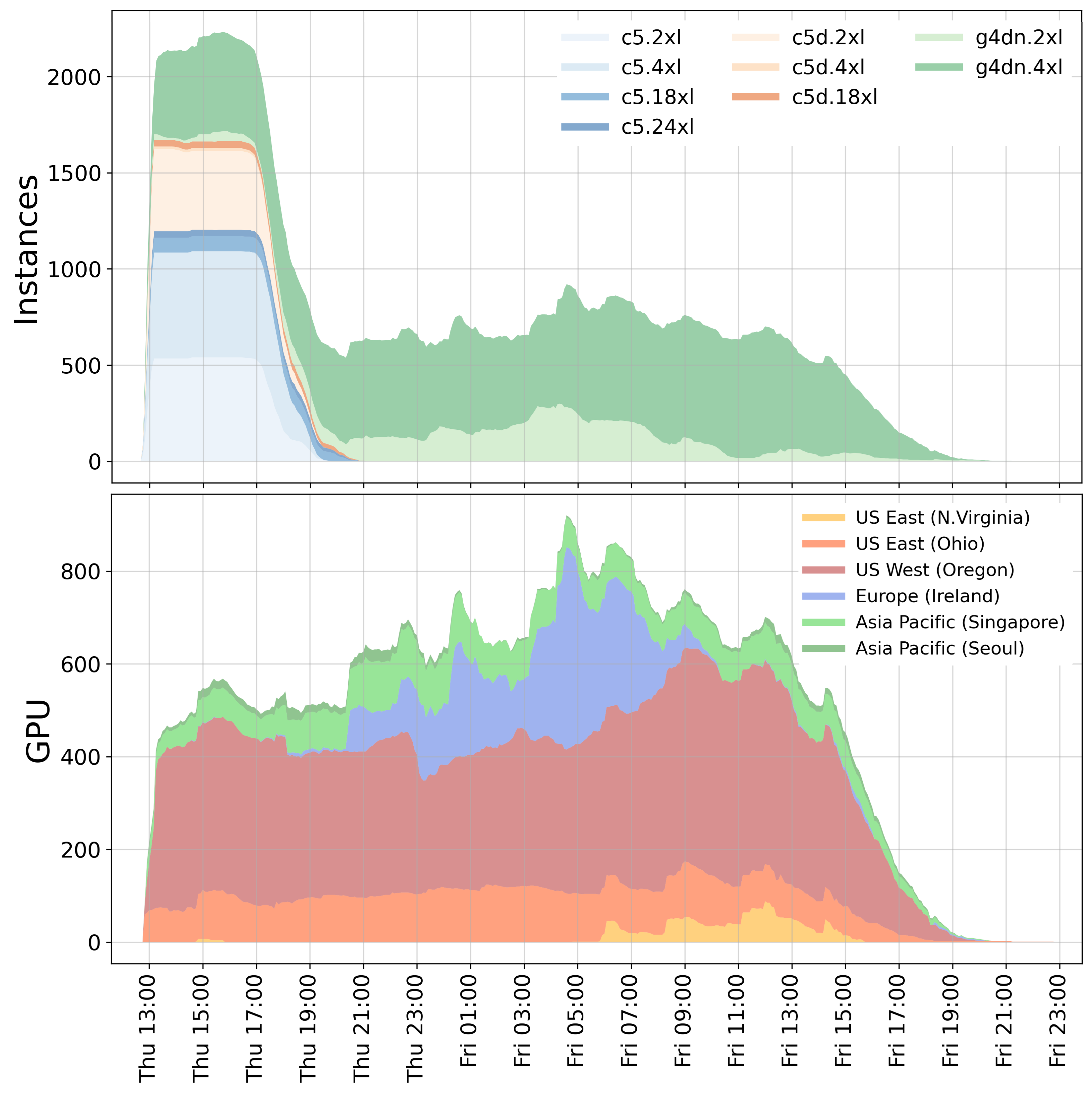}\\
\caption{\textbf{Usage of global compute resources for the second ligand screening study aimed to optimize cost-efficiency.}
Top: Allocated instance types over time,
bottom: GPU instances allocated in the different regions.}
\label{fig:run2}
\end{figure}

The setup was as in the first study, however we further excluded instances with low cost-efficiency:
most notably, we ran all the protein systems on cost-efficient GPU instances.
The acquired resources over time for the second study are shown in Fig.~\ref{fig:run2}.

The vCPU usage peaked at slightly above 35,000 vCPUs at two hours into the second run (not shown),
with on average 500 GPU instances in use over 26 hours.
About six hours after the submission of the ensemble the small ligand in water systems were finished 
(blue and orange areas in Fig.~\ref{fig:run2} top).
As our benchmarks on \texttt{c5.2xl} estimated a runtime of about five hours per system,
we conclude that there were enough \texttt{c5} Spot instances available to run each of the 3,492 ligand jobs concurrently.

GPU instances are however running over a time span of about 30 hours altogether,
as apparently not enough \texttt{g4dn} Spot capacity was available to run all 3,492 GPU jobs concurrently.
From the lower panel of Fig.~\ref{fig:run2} we see that at the time of submission, 
there was only \texttt{g4dn} capacity available in four regions, 
whereas the Ireland (blue) and North Virginia (yellow) regions provided \texttt{g4dn} instances only after several hours into the run.
The large differences across regions underline that a multi-region approach is necessary
for decent job throughput when limiting oneself to few instance types only.

The resulting costs of our second study were about 16 \$ per FE difference,
thus only about a third of what was achieved in the first study and in line
with what is expected from the benchmarks on \texttt{g4dn} instances (Fig.~\ref{fig:optimalConfig}).

Both high throughput ligand screening studies illustrate the flexibility of cloud computing for MD-based investigations:
AWS can be used to scale up the computations to the extent of a large HPC facility, but can also be used in a cost-efficient manner akin 
to a smaller in-house cluster.
When aiming to minimize the time-to-solution, the 19,872 calculation jobs were finished in $\approx$2 days.
This compares well to the timing in the recent report, where the Tier 2 Max Planck Supercomputer Raven (interim version, 
480 Intel Xeon Cascade Lake-AP nodes with 96 cores, 192 threads) performed calculations of the same dataset in $\approx$3 days.\cite{gapsys2022raven}
The cost-efficient usage of the cloud resources allowed reaching the cost of 16 \$ for a free energy calculation.
Cost-efficiency could be further optimized by also running the ligand in water simulations on the \texttt{g4dn} GPU instances
(instead of using \texttt{c5} instances),
which would result in a cost of 14 \$ per free energy difference, although \texttt{g4dn} capacity may then
not be sufficient to run all simulations at once.
In comparison to a \gromacs optimized in-house cluster of Intel E5-2630v4 10-core nodes and NVIDIA RTX 2080 GPU,
this cost would be $\approx$8.5 \$,
in agreement with the estimates of relative costs for a compute node analyzed in Fig.~\ref{fig:compareTCO}.

\section{Summarizing discussion}
Cloud computing has the potential to transform large-scale simulation projects. 
To date, computationally intensive projects,
when assigned to busy on-premises clusters with limited computing capacity,
may need weeks or months to be completed.
In the cloud, though, the required processing power can be distributed among numerous compute centres around the globe.
With the removal of the bottleneck of limited computing capacity,
jobs that are independent of each other can run simultaneously in a high throughput manner,
thus reducing the time-to-solution to the runtime of the longest simulation of the ensemble.
Such use cases that require very high peak performance over a short period of time can easily be met by cloud computing,
while installing and operating a sufficiently large local cluster would be neither cost-effective nor feasible.

For the use case of MD-based high throughput ligand screening we established a HyperBatch-based workflow
that allows to complete large-scale projects that would run for weeks on an on-premises cluster
within 48 hours or less in the cloud.
Shortly after submitting 19,872 jobs,
we acquired about 140,000 compute cores and 3,000 GPUs in multiple regions around the globe.
We demonstrated that the costs associated with such projects can be reduced about nine-fold 
compared to a na{\"i}ve implementation:
A job checkpoint-restart mechanism allowed to use Spot instances instead of on-demand instances,
which accounts for a threefold reduced price.
Putting the benchmarked application performance in relation to the instance costs
allowed to select from a huge variety of available instance types the most cost-efficient ones only,
thereby reducing the price tag by another factor of three,
albeit at the cost of a longer time-to-solution.

Whereas HyperBatch is geared towards speeding up HTC projects,
we also investigated HPC strong scaling scenarios with a cloud-based HPC cluster.
Cluster installation via ParallelCluster and software installation via Spack provided a straightforward
and reproducible setup.
Due to the possibility to automatically scale up (and down) the number of cluster nodes
depending on whether there are jobs in the queue,
there is virtually no waiting time for jobs in the queue.
The breadth of readily available hardware that includes several architectures (Intel, AMD, ARM) in
various sizes (regarding core count), accelerators like GPUs, and high-performance network if wanted,
allows to always pick the optimal hardware for the job at hand,
in terms of a short time-to-solution or a high cost-efficiency.
For \gromacs, we found that \texttt{g4dn} instances offer the highest performance to price ratio,
whereas instances with the fastest interconnect (\texttt{c6i.32xl} and \texttt{c5n.18xl}) showed the best parallel scaling
on up to 64 instances using 8,192 vCPUs altogether for the largest benchmark system.

So how did, overall, cloud computing compare to a local cluster for our realistic test cases?
For many cases, the extra flexibility offered by the cloud will certainly come at a cost 
higher than that of a local compute cluster. 
However, as our study shows, by aggressively tuning both alternatives towards cost-efficiency
we are approaching a break even point, and the costs of cloud computing and on-premises computing become similar. 
In fact, an on-premises consumer-GPU cluster tailored towards \gromacs
produces MD trajectory at about 2/3 of the costs of Spot GPU instances
with similar performance.

We note that this outcome is also due to the fact that very specialised single software, \gromacs, was used;
in contrast, if a wider variety of software has to run, the nodes can probably not be tuned that
much and therefore will be less cost-efficient for a particular application.
Just the use of a professional GPU instead of a consumer GPU will result in trajectory
costs significantly higher than what can be achieved on an optimal Spot instance.


\section{Conclusions}

Cloud computing has traditionally been way more expensive than an on-premises cluster 
for the case that continuous long term computer performance is required.
Here we have shown that this has changed, 
at least for the specialised, yet highly important application of drug design.
We are now at a break even, where the costs are the same, 
maintaining the great benefit of cloud computing to offer enormous flexibility and, 
if required, extremely short production times. 
We consider this a critical milestone for MD-based high throughput computational drug design.


\section*{Data and Software Availability}
The input files for the benchmarks can be downloaded from \url{https://www.mpinat.mpg.de/grubmueller/bench}.
A guide to build GROMACS on AWS is available here: \url{https://gromacs-on-pcluster.workshop.aws}.

\section*{Acknowledgments}
Compute time for all cloud-based simulations of this study was generously provided by AWS public sector.
Many thanks to Torsten Bloth, Stephen Sachs, Cristian M\u{a}gheru\c{s}an-Stanciu, 
Bruno Silva, Agata Jablonka, and Johannes Schulz for advice and support throughout the project.
Simulation input preparation and output analysis was done on compute clusters of the Max Planck Society.
The work was supported by the BioExcel CoE (www.bioexcel.eu), a project funded by the European Union contracts H2020-INFRAEDI-02-2018-823830.

\providecommand{\latin}[1]{#1}
\makeatletter
\providecommand{\doi}
  {\begingroup\let\do\@makeother\dospecials
  \catcode`\{=1 \catcode`\}=2 \doi@aux}
\providecommand{\doi@aux}[1]{\endgroup\texttt{#1}}
\makeatother
\providecommand*\mcitethebibliography{\thebibliography}
\csname @ifundefined\endcsname{endmcitethebibliography}
  {\let\endmcitethebibliography\endthebibliography}{}

\end{document}